\documentclass[
nofootinbib,
 amsmath,amssymb,
 aps,
 prd,
 twocolumn
]{revtex4-1}

\usepackage{color}
\usepackage{ulem}
\usepackage[subnum]{cases}
\renewcommand\sout{\bgroup \color{blue} \ULdepth=-.5ex \ULset}

\newcommand{\cU}{\mathcal{U}}

\newcommand{\lbar}{\bar{\ell}}

\usepackage{graphicx}
\usepackage{dcolumn}
\usepackage{bm}
\usepackage{hyperref}

\begin{document}

\preprint{APS/123-QED}

\title{Deconfinement in the presence of a strong magnetic field}

\author{Pok Man Lo}
\affiliation{Institute of Theoretical Physics, University of Wroclaw,
PL-50204 Wroc\l aw, Poland}
\author{Micha\l {} Szyma\'nski}
\affiliation{Institute of Theoretical Physics, University of Wroclaw,
PL-50204 Wroc\l aw, Poland}
\author{Krzysztof Redlich}
\affiliation{Institute of Theoretical Physics, University of Wroclaw,
PL-50204 Wroc\l aw, Poland}
\affiliation{Theoretical Physics Department, CERN, CH-1211 Gen\`eve 23, Switzerland}
\author{Chihiro Sasaki}
\affiliation{Institute of Theoretical Physics, University of Wroclaw,
PL-50204 Wroc\l aw, Poland}

\date{\today}

\begin{abstract}

We study the impact of a finite magnetic field on the deconfinement phase transition for heavy quarks by computing the fluctuations of the Polyakov loops.
It is demonstrated that the explicit Z(3) breaking field increases with the magnetic field, leading to a decrease in the (pseudo) critical temperatures
and a shrinking first-order region in the phase diagram.
Phenomenological equations which capture the behaviors of the Z(3) breaking field at strong and weak magnetic fields for massive and massless quarks are given.
Lastly,  we explore the case of dynamical light quarks, and demonstrate how an improved constituent quark mass function
can enforce the correct magnetic field dependence of the deconfinement temperature in an effective model, as observed in  Lattice QCD calculations.

\end{abstract}

\maketitle

\section{Introduction}
\label{sec1}

Magnetic fields provide an interesting handle to probe QCD properties under extreme conditions~\cite{DElia:2012ems, Kharzeev:2012ph, Shovkovy:2012zn, Andersen:2014xxa, Miransky:2015ava}.
Investigating its influences in the phase diagram of strongly interacting matter is important for understanding the physics of
noncentral heavy-ion collisions~\cite{Skokov:2009qp, Voronyuk:2011jd, Bzdak:2011yy, Deng:2012pc, Tuchin:2013ie},
the bulk properties of high-field neutron stars~\cite{Duncan:1992hi, Ferrer:2012wa} and possibly the early Universe~\cite{Grasso:2000wj}.

Lattice simulations~\cite{DElia:2011koc,Bali:2011qj,Bali:2012zg,Bruckmann:2013oba,Bornyakov:2013eya,Bali:2014kia,Endrodi:2015oba,DElia:2018xwo,Endrodi:2019zrl}
performed for light quarks predict that
the chiral condensate in vacuum is enhanced by the presence of a magnetic field $B$, a phenomenon known as magnetic catalysis~\cite{DElia:2011koc,Bali:2012zg,Bruckmann:2013oba,Bornyakov:2013eya},
and that the critical temperature for the chiral phase transition is decreasing with $B$, i.e. inverse magnetic catalysis~\cite{Bali:2011qj,Bali:2012zg,Bornyakov:2013eya,Endrodi:2015oba,DElia:2018xwo,Endrodi:2019zrl}. It is also found that the manifestation of these phenomena depends strongly on the pion mass~\cite{DElia:2018xwo,Endrodi:2019zrl}.

While most effective chiral models (e.g. the NJL model) can capture the effect of magnetic catalysis,
they tend to predict the opposite trend in the $B$-field dependence of the chiral transition temperature~\cite{Fraga:2012rr, Fraga:2013ova}.
To accommodate a decreasing transition temperature with $B$ one usually needs to introduce additional parameter dependence in the effective potentials,
e.g. a $B$-dependent coupling~\cite{Farias:2014eca, Ferrer:2014qka}, and imposing a specific treatment of the vacuum (and thermal) fluctuations. 
See also Refs~\cite{Pagura:2016pwr, Dumm:2018oop, GomezDumm:2017jij} for a nonlocal extension of the PNJL model. This is far from ideal as it points to missing interactions (e.g. backreactions, higher order terms, etc.)
that are unaccounted for in the original model, demanding a more careful explicit treatment~\cite{Braun:2007bx, Schaefer:2007pw, Reinosa:2014ooa, Reinosa:2015oua, Fukushima:2012qa, Fukushima:2017csk}.~\footnote{Even at $B=0$, a simple implementation of the PNJL model already leads to a substantially higher transition temperature than LQCD. Implementing a running $T_d$~\cite{Fraga:2013ova, Braun:2007bx} may improve the situation, but the curvatures of the potential and various fluctuations remain to be tested.}

An important diagnostic test for the correct form of the potential (rather than a parameter change) is to examine higher order fluctuations: as a first step we shall study the Polyakov loop fluctuations in the presence of magnetic field.
Ratios of these fluctuations are excellent probes of deconfinement: In a pure gauge theory, they exhibit a jump at the transition temperature~\cite{Lo:2013etb,Lo:2013hla}, with well-defined low temperature limits deducible from general theoretical constraints and the Z(3) symmetry. Even in full QCD, they provide a measure for the strength of explicit Z(3) symmetry breaking field induced by the light fermions, and show less renormalization scheme dependence than the Polyakov loop~\cite{Bazavov:2016uvm,Lo:2018wdo}.

\begin{figure*}[ht]
\centering
\includegraphics[width=0.49\linewidth]{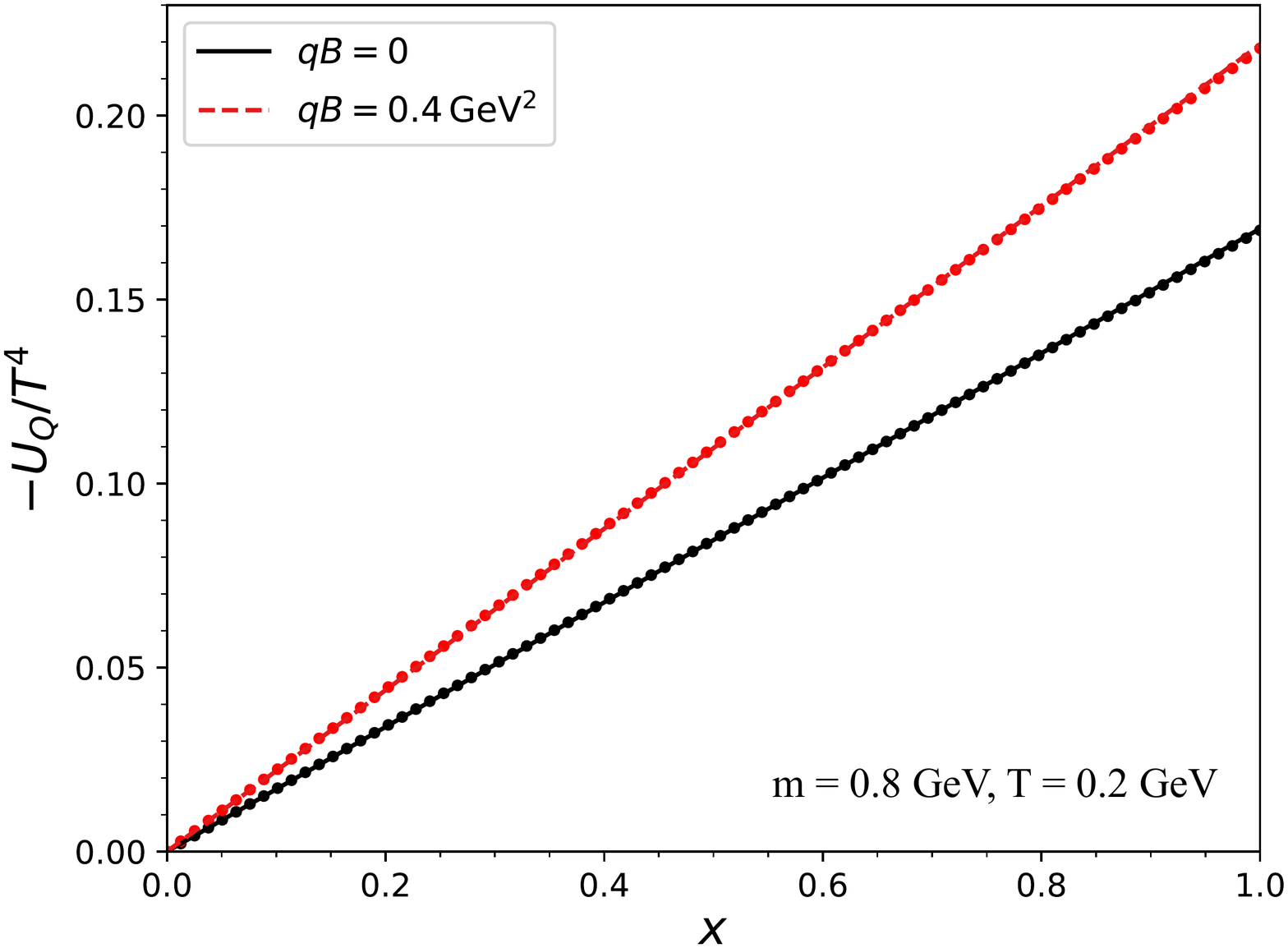}
\includegraphics[width=0.49\linewidth]{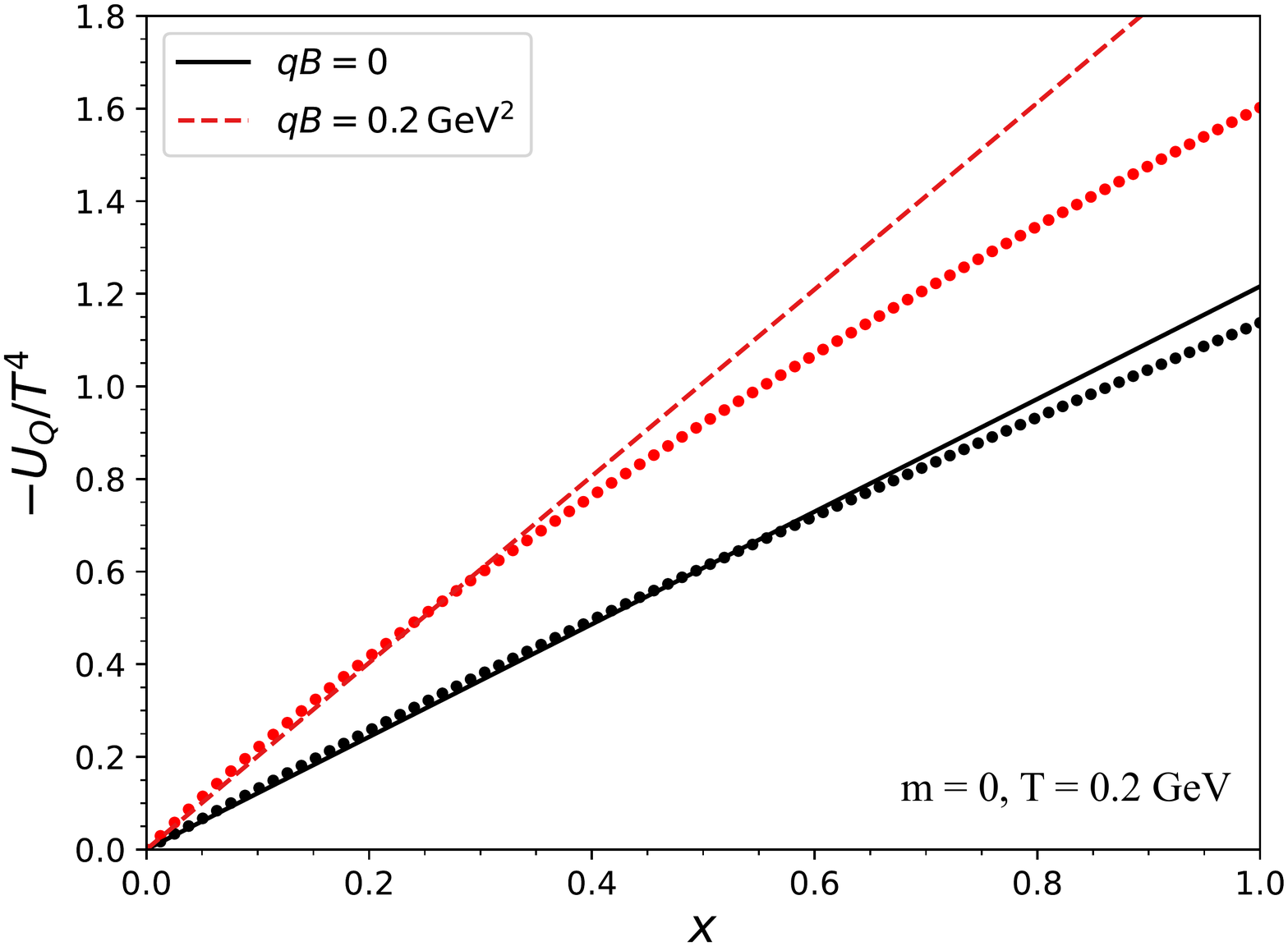}
	\caption{Comparison between the full quark potential \eqref{eq:UQfull}
(points), and its linear approximations \eqref{eq:lin_approx} (lines), versus the real part of the Polyakov loop field for massive (left) and massless (right) quarks. The effect of an external magnetic field on the potential is also shown.
For quark masses above $0.8$ GeV, the regime explored in this work, the linear approximation provides an adequate description of the full potential.
}
\label{fig1}
\end{figure*}

To describe these fluctuations in an effective model, not only the location, but also the curvatures around the minima of the Polyakov loop potential have to be adjusted. 
Similar concept applies to the matrix models~\cite{Meisinger:2001cq, Dumitru:2012fw, Dumitru:2013xna, Kashiwa:2012wa, Kashiwa:2013rm}, where variances, in addition to mean values, of the distribution of the eigenvalues of the thermal Wilson line, should be examined.
The curvatures dictate how reluctant the system is to deviate from the equilibrium position in the presence of an external disturbance. Lattice calculations show that the Polyakov loop generally increases with the magnetic field strength~\cite{Bali:2011qj, Bruckmann:2013oba,Bornyakov:2013eya,Endrodi:2015oba, DElia:2018xwo, Endrodi:2019zrl}. Also, the resulting pseudocritical temperature of deconfinement decreases with the magnetic field~\cite{Endrodi:2015oba,DElia:2018xwo,Endrodi:2019zrl}, observed for both light and heavy pions~\cite{DElia:2018xwo,Endrodi:2019zrl}.
On the other hand, the response of the fluctuations to an external magnetic field has not yet been studied, and one of the goals of the current paper is to fill this gap.

In this work we extend the model introduced in Ref.~\cite{Lo:2014vba}
to a finite magnetic field and study its effect on deconfinement in a system with heavy quarks.
The study of deconfinement of heavy quarks is interesting in its own right as it relates to other issues such as
the modification of the heavy quark potentials in strong fields~\cite{Bonati:2015dka} and color screening~\cite{Bonati:2017uvz, Rucci:2019hcd}.
In addition, restricting to the heavy quarks allows us to avoid the complications from the chiral transitions.
This serves as a foundation for setting up a reliable effective gluon potential to further assess the delicate interplay
between light quarks and gluons at finite magnetic field.

In our approach the effect of dynamical quarks is modeled by a linear Z(3) breaking term, coupled to the Polyakov loop.
This term becomes Landau-quantized when finite magnetic field is present. With such the model setup we explore the Polyakov loop as well as its fluctuations.
We shall explore in details how an external magnetic field would enhance the explicit Z(3) breaking, and could eventually change a first order phase transition into a crossover.
Moreover, the deconfinement temperature decreases with the magnetic field. We also discuss the phase diagram in the $(m_l,m_s)$ plane and find that the external magnetic field shrinks the first order region.

Lastly we explore the case of dynamical light quarks within the PNJL model.
The well-known problem of an increasing deconfinement transition temperature with $B$ can easily be understood in the current model.
We shall also demonstrate,  how the use of an improved constituent quark mass function can produce instead the trend of a decreasing deconfinement transition temperature with $B$.

\section{Polyakov loop potential at finite magnetic field }
\label{sec2}

\begin{figure*}[ht]
\centering
\includegraphics[width=0.49\linewidth]{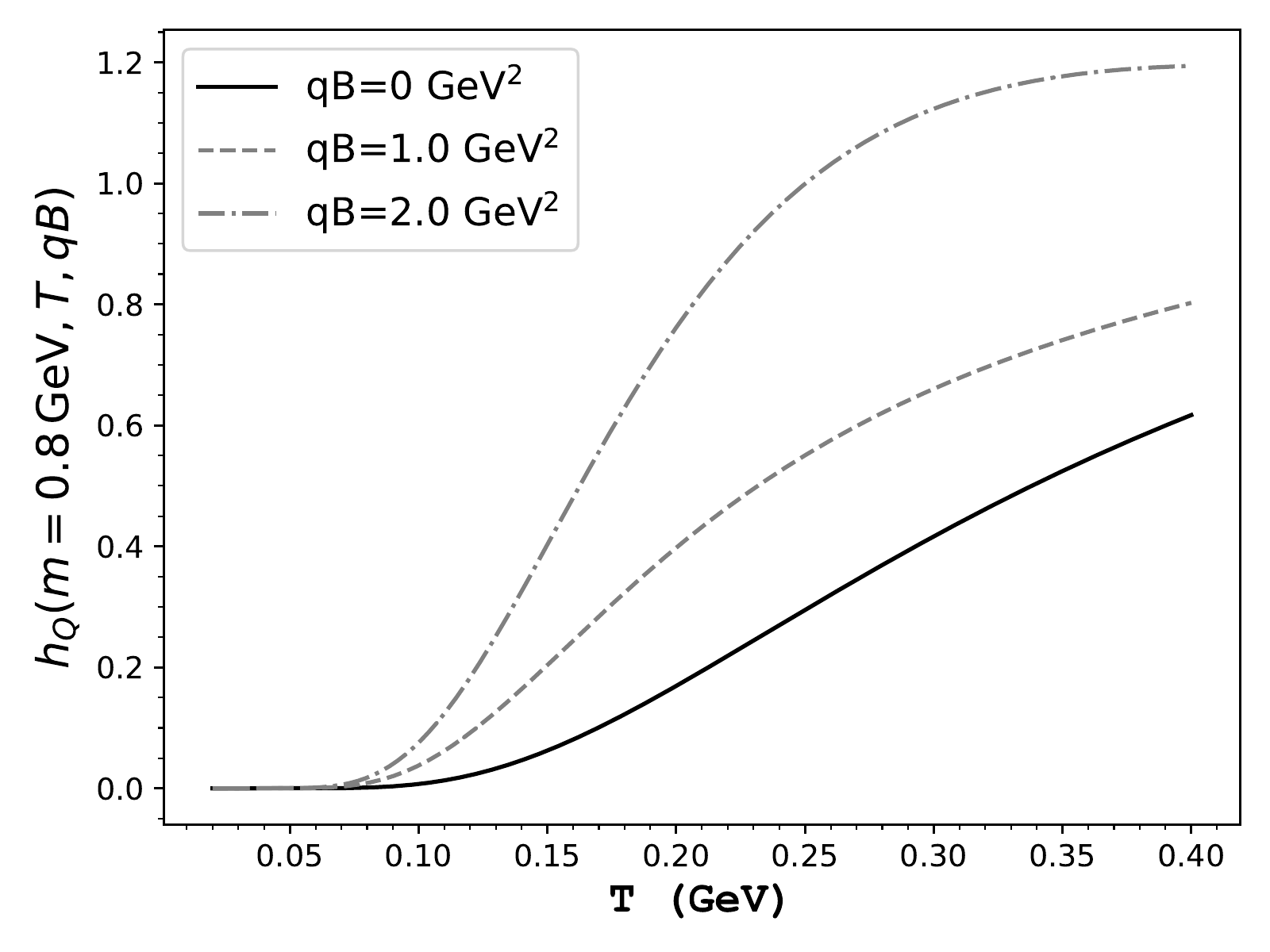}
\includegraphics[width=0.49\linewidth]{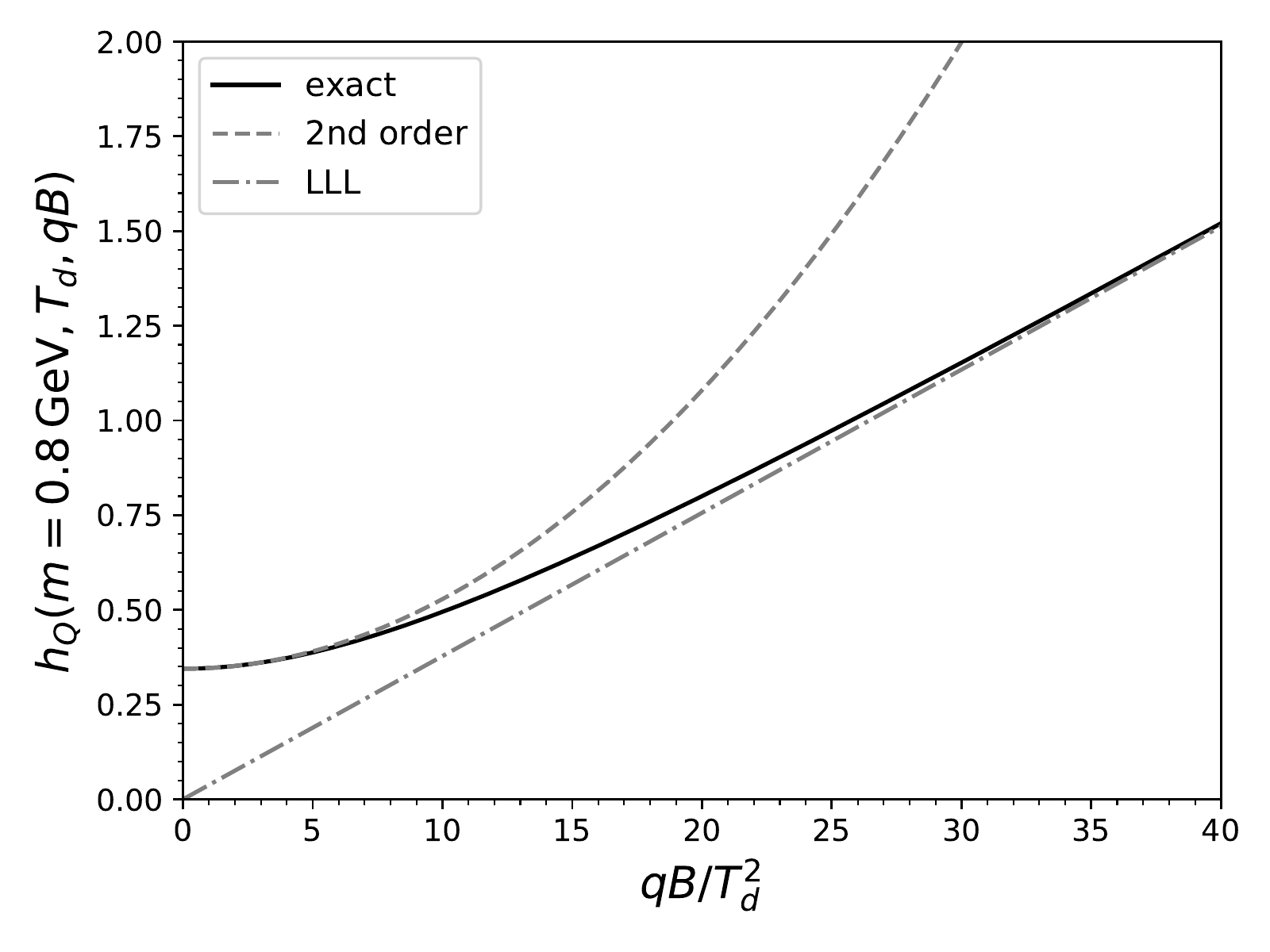}
	\caption{Left: the Z(3) symmetry breaking field $h_Q^B$ \eqref{eq:hfullB} versus $T$ at various magnetic field strengths.
	Right: the $B$-field dependence of $h_Q^B$ at fixed $T_d = 0.27$ GeV.
	The exact numerical result of $h_Q^B$ smoothly interpolates between the weak (2nd order) and strong (LLL) field limits in Eq. \eqref{eq:key1}.}
\label{fig2}
\end{figure*}

\subsection{Robustness of the linear approximation}

We model the effective Polyakov loop potential as follows:

\begin{align}
    \cU=\cU_G+\cU_{Q},
\end{align}
where $\cU_G$ is a pure glue potential and $\cU_{Q}$ describes the explicit Z(3) symmetry breaking due to dynamical quarks.
The Z(3) symmetric pure glue potential $\cU_G$ of choice is from Ref~\cite{Lo:2013hla}:

\begin{align}
	\label{eq:UG}
	\frac{\cU_G}{T^4} &= -\frac{A}{2}\lbar \ell+B\ln M_H(\ell,\lbar)\nonumber\\
&+\frac{C}{2}(\ell^3+\lbar^3)+D(\lbar \ell)^2
\end{align}
where
\begin{align}
\ell&=x+iy\,,\\
\lbar&=x-iy\,
\end{align}
are the Polyakov loop and its conjugate. $M_H(\ell,\lbar)$ is the SU(3) Haar measure,
\begin{align}
M_H(\ell,\lbar)=1-6\lbar \ell+4(\ell^3+\lbar^3)-3(\lbar \ell)^2,
\end{align}
and $A$, $B$, $C$ and $D$ are temperature-dependent model parameters determined from the lattice QCD (LQCD) results on the pressure, the Polyakov loop and its susceptibilities in a pure gauge theory~\cite{Lo:2013hla}.

The $\cU_Q$ potential describes the coupling between quarks and the Polyakov loop. To one-loop order the expression reads~\cite{Lo:2014vba, Kashiwa:2012wa}

\begin{align}
	\label{eq:UQfull}
	\frac{\cU_Q}{T^4}&= -\frac{2}{T^3} \sum_{f=u,d,s} \int \frac{d^3 {p}}{(2 \pi)^3} \, [ \ln  g^+ + \ln g^- ]
\end{align}
with
\begin{align}
        g^+ &= (1 + 3 \, \ell \, e^{-\beta E} + 3 \,  \lbar \, e^{-2 \beta E} + e^{-3 \beta E})  \nonumber \\
        g^- &= (1 + 3 \, \lbar \, e^{-\beta E} + 3 \, {\ell} \, e^{-2 \beta E} + e^{-3 \beta E}),
\end{align}
where the sum runs over different flavors (up, down and strange) and $E = \sqrt{p^2 +m_f^2}$.

In previous studies, we find that the potential $\cU_Q$ can be approximated fairly accurately by the leading term in $\ell$, i.e.
\begin{align}
	\label{eq:lin_approx}
	\frac{\cU_Q}{T^4} \approx  - \sum\limits_{f=u,d,s} h_Q(m_f,T) \times \frac{1}{2} \, (\ell + \lbar),
\end{align}
where the leading constant term $\cU_Q(\ell,\lbar=0)$ is dropped as it is irrelevant for computing the Polyakov loop and its fluctuations.
The Z(3) breaking strength $h_Q$ is given by~\cite{Kashiwa:2012wa,Lo:2014vba}

\begin{align}
	\label{eq:h0}
	\begin{split}
		h_Q(m, T) &= \frac{12}{T^3} \, \int \frac{d^3 p}{(2 \pi)^3} \, e^{-\beta E(p)} \\
			  &= \frac{6}{\pi^2} \, (\frac{m}{T})^2 \, K_2(m/T).
	\end{split}
\end{align}
where $K_2(x)$ is the modified Bessel function of the second kind.
The pure gauge limit is recovered  at $m \rightarrow \infty$, giving $h_Q \rightarrow 0$.
The opposite limit, $m \rightarrow 0$, is also of interest, where we find $h_Q \rightarrow 12/\pi^2$.
This means that while the explicit Z(3) breaking field gradually increases as quark masses are decreased,
it does not do so indefinitely but saturates at a maximum value.
This should be contrasted with the case of chiral symmetry breaking, where the quark mass $m$ serves as the linear breaking field, and can increase without limit.

Unfortunately the linear breaking strength for Z(3) is not directly measured in most LQCD studies.
A crude estimate~\cite{Lo:2018wdo} based on LQCD results on the ratios of Polyakov loop susceptibilities suggests a larger Z(3) breaking strength than the prediction from the  PNJL model. Nevertheless, the issue is far from settled as the analysis is still marred by the unsolved problem of the proper renormalization of the Polyakov loop and its susceptibilities. The study of the quark mass and the magnetic field dependencies of the Z(3) breaking field may provide some hints for tackling the problem.

The linear approximation made in Eq.~\eqref{eq:lin_approx} is expected to work for heavy quarks.
We now examine the efficacy of the scheme for lower (and even vanishing) quark masses.
A direct comparison of the full potential $\cU_Q$ and its linear approximation is shown on Fig.~\ref{fig1}.
Here, the temperature is fixed at $T = 0.2$ GeV and we show the case of $m_q=0.8$ GeV (left) and a massless quark (right).
We see that the linear approximation provides an excellent description of the full potential for quark masses $\geq 0.8$ GeV:
the two results are almost indistinguishable.
The scheme remains fairly robust even in the worst case scenario of a massless quark, giving only small differences at large values of Polyakov loop $x \geq 0.7$.

As a preview we also show in Fig.~\ref{fig1} the influence of a finite magnetic field $B$ on the Polyakov loop potential.
The main effect is that the Z(3) breaking strength, corresponding to the slope of the Z(3) breaking potential, is enhanced.
We see that for massive quarks even at a moderate $qB=0.4 \, {\rm GeV}^2$ the linear scheme still provides a very good approximation.
In fact for the range of parameters explored in this work, the difference between the full and the linear approximation is negligible.
Nevertheless it is not the case for light or massless quarks. For example, we see substantial deviation for massless quark at $qB = 0.2 \, {\rm GeV}^2$, which is a typical field strength when studying chiral transitions~\cite{Andersen:2014xxa}.

The versatility of the linear approximation for gaining intuitive understanding shall become obvious in the discussion.
In any case, a direct numerical computation of the full potential $\cU_Q$ is rather straightforward and is always an option for quantitative study.
In the next sections, a more detailed analysis of the functional dependence on the field strength $B$ and its effects on the fluctuation observables will be explored.

\subsection{ Z(3) breaking strength at finite $B$ }


In a constant and homogeneous magnetic field background the motion of charged particles undergoes the Landau quantization in the transverse plane. Consequently, the dispersion relation is modified and takes the following form for spin $1/2$ particles,
\begin{align}
E_{f,k,\sigma}^{\ 2}=m_f^2+p_z^2+ (2k+1-\sigma)\vert q_f B\vert\,,
\end{align}
where the subsequent Landau levels are quantified by $k=0,1,2...$ and $\sigma=0, 1$ (the spin projection on the $\vec{B}$ axis). The sum over states is modified accordingly:
\begin{align}
2\int\frac{d^3p}{(2\pi)^3}\rightarrow  \frac{\vert qB\vert}{2\pi}\sum\limits_{\sigma=\pm 1}\sum\limits_{k=0}^\infty\int\limits_{-\infty}^\infty \frac{dp_z}{2\pi}\,,
\label{eq:landau}
\end{align}
where the summation runs over Landau levels and the factor $\vert qB\vert/(2\pi)$ accounts for the planar density of each Landau level. Consequently, the explicit Z(3) breaking term becomes a function of the magnetic field,

\begin{align}
	h_{\rm tot} = \sum\limits_{f=u,d,s}h_Q^B(m_f, T, q_f B),
\end{align}
where $h_Q^B(m, T, qB)$ is obtained by applying the prescription \eqref{eq:landau} to Eq. \eqref{eq:h0},
\begin{align}
\begin{split}
h_Q^B(m,T,qB)=\frac{3\vert qB\vert}{2\pi^2T^3} \sum\limits_{\sigma=\pm 1}\sum\limits_{k=0}^\infty\int\limits_{-\infty}^\infty dp_z e^{-E_{f,k,\sigma}/T}\,.
\end{split}
\end{align}
The integration over $dp_z$ can be performed analytically, which leads to

\begin{align}
\label{eq:hfullB}
h_Q^B(m,T,qB)=\frac{3\vert qB\vert}{\pi^2T^3} \sum\limits_{\sigma=\pm 1}\sum\limits_{k=0}^\infty M_{k,\sigma} K_1\left(M_{k,\sigma}/T\right)\,,
\end{align}
where
\begin{align}
	M_{k,\sigma}= \sqrt{ m^2+(2k+1-\sigma)\vert qB\vert}.
\end{align}

The main effect of a finite magnetic field on the Polyakov loop is to increase the Z(3) breaking strength.
Fig. \ref{fig2} (left) shows the typical behavior of the Z(3) breaking field at finite $B$.
For a given $B$, the increase is more rapid for lower values of $m/T$.

The numerical computation of Eq.~\eqref{eq:hfullB},
or even the full $\cU_Q$ using the prescription \eqref{eq:landau},
is straightforward and has been explored in previous works. See for example Refs.~\cite{Fraga:2012rr, Andersen:2014xxa}
and also Fig. \ref{fig1}.
On the other hand, for an intuitive understanding of the magnetic field dependence
we shall present both the weak and strong field limits of the linear breaking term, for massive and massless quarks.

We start with the case of massive quarks. Since $h_Q^B$ is dimensionless,
it can be expressed in terms of the dimensionless combinations of $m/T$ and $\vert qB\vert/T^2$. The weak and strong field limits read:

\begin{widetext}
	\begin{align}
	\label{eq:key1}
	h_Q^B(m,T,qB) \approx \begin{cases} \frac{6}{\pi^2}( m/T)^2 K_2(m/T)+\frac{1}{2\pi^2}K_0\left(m/T\right)\left(\frac{\vert qB\vert}{T^2}\right)^2,\ & \vert qB\vert\ll m^2,\ T^2 \\
		\frac{3}{\pi^2} \, (m/T) \,  K_1\left(m/T\right) \, \frac{\vert qB \vert}{T^2},\ & \vert qB\vert\gg m^2,\ T^2
	\end{cases}.
	\end{align}
\end{widetext}
The result is obtained from an asymptotic series expansion of Eq.~\eqref{eq:hfullB}. A detailed derivation, as well as a discussion of higher order terms, are presented in the Appendix~\ref{app1}. The first term in Eq. \eqref{eq:key1} coincides with Eq. \eqref{eq:h0}, while the corrections start at the quadratic order, i.e. there is no linear order correction in $B$ in the weak field limit. Due to the coefficient $K_0(m/T)$, the response to $B$ is more suppressed for heavier quarks.

When magnetic field becomes strong, the Z(3) breaking field is dominated by the lowest Landau level (LLL), in which

\begin{align}
\label{eq:LLL}
h_{LLL}(m,T,qB)=\frac{3}{\pi^2} \, (m/T) \,  K_1\left(m/T\right) \, \frac{\vert qB \vert}{T^2},
\end{align}
while the higher levels are exponentially suppressed.
Note, that in the strong field limit the Z(3) breaking field becomes linearly dependent on $B$.
The exact numerical result of Eq.~\eqref{eq:hfullB} smoothly interpolates between the two limits. See Fig. \ref{fig2} (right).

It is also of interest to study the case of massless quarks.
It turns out that the corresponding weak field limit cannot be obtained directly from the $m \rightarrow 0$ limit of Eq. \eqref{eq:key1}. Instead we have the following result:

\begin{widetext}
\begin{align}
\label{eq:key2}
	h_Q^B(m=0, T, qB) \approx \begin{cases} \frac{12}{\pi^2}+\frac{1 -2\gamma_E  +2\ln 2-12\zeta'(-1)}{4\pi^2} \left(\frac{|qB|}{T^2}\right)^2 -\frac{1}{4\pi^2} \left(\frac{|qB|}{T^2}\right)^2 \ln\left(\frac{2|qB|}{T^2}\right)  -\frac{3\zeta'(-2)}{4\pi^2}\left(\frac{|qB|}{T^2}\right)^3, \ & \vert qB \vert \ll T^2  \\
	\frac{3}{\pi^2} \, \frac{\vert qB\vert}{T^2},\ & \vert qB\vert \gg T^2
	\end{cases}.
\end{align}
\end{widetext}
For weak fields, beyond the leading quadratic correction, there exist the peculiar odd-power and logarithmic terms.
The computational details can be found in the Appendix \ref{app1}.
Lastly, the strong field limit is dictated by the LLL, and similar to the massive case, the Z(3) breaking field becomes linearly dependent on $B$.

To summarize the effects of a finite magnetic field to the Z(3) breaking fields: First, it tends to increase the explicit Z(3) breaking strength. Second,
the corrections start with quadratic order in $qB$ for weak fields, and are dominated by the LLL at strong fields. The latter gives a linear dependence.
Lastly, the increase is more rapid for light than heavy quarks.
\begin{figure*}[!ht]
\centering
\includegraphics[width=0.49\linewidth]{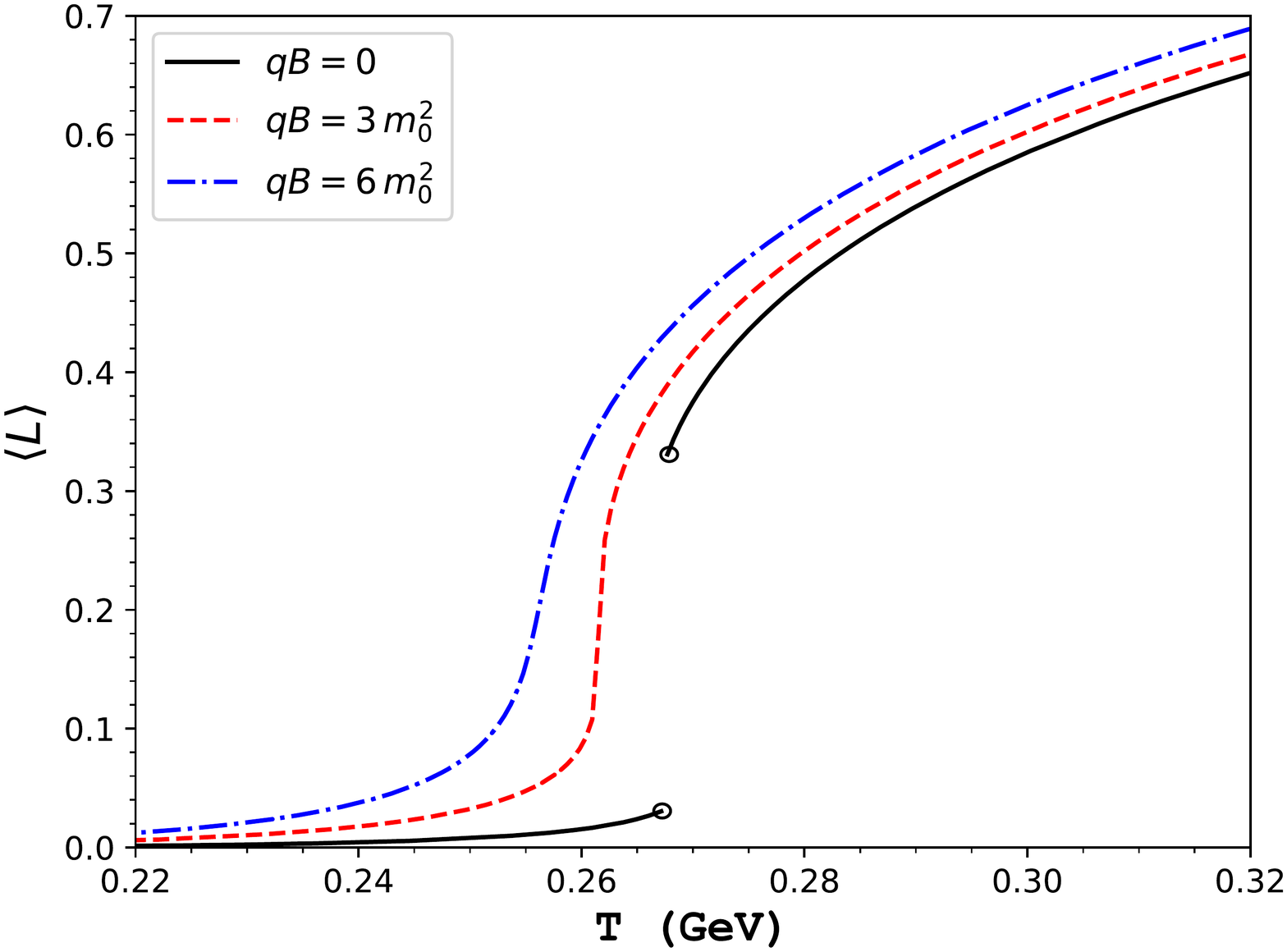}
\includegraphics[width=0.49\linewidth]{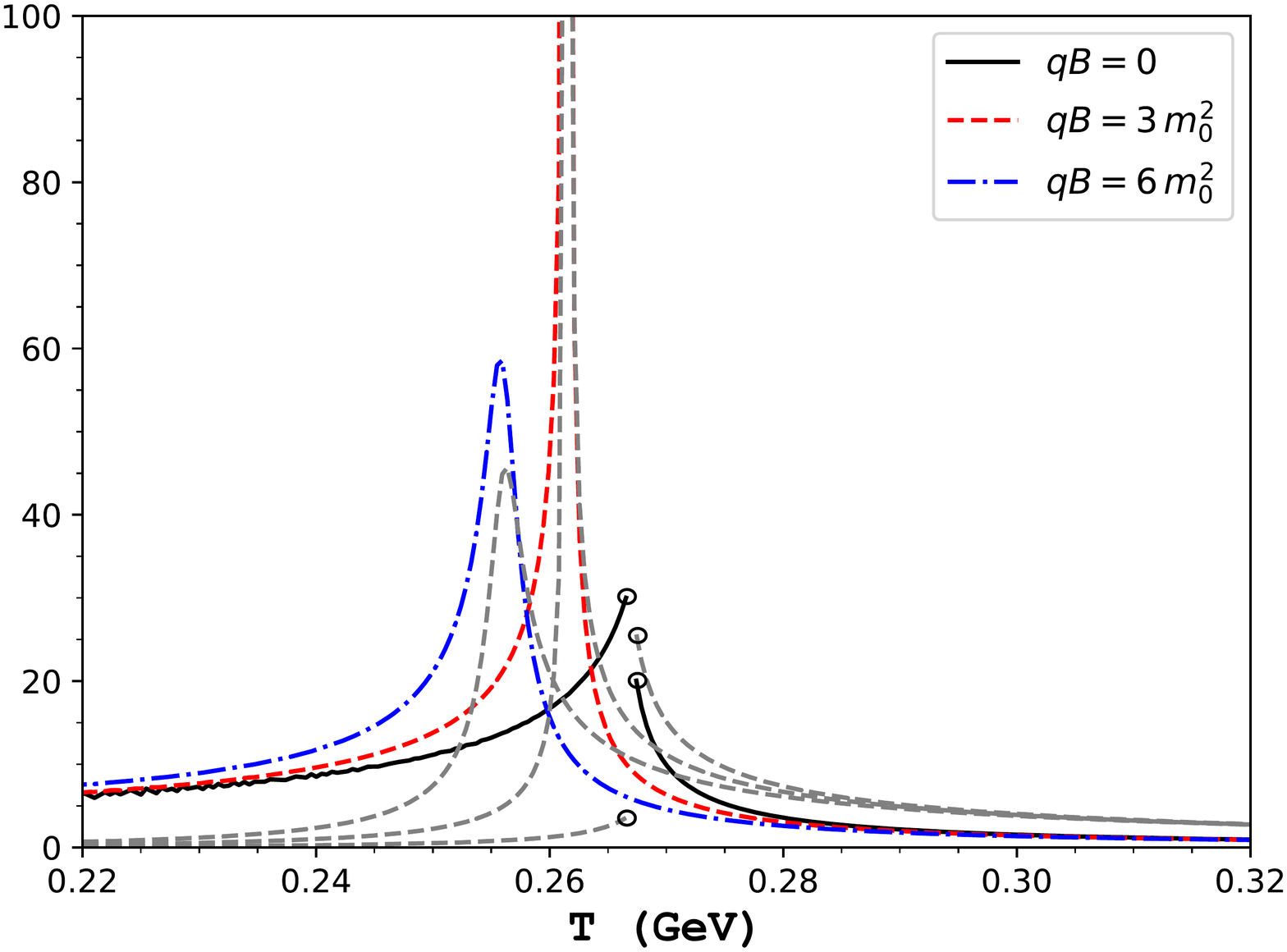}
	\caption{The Polyakov loop expectation values (left) and the single quark entropy $S_Q$ (right) at $qB=(0,3,6) \, m_0^2$ for a single strange-quark flavor system at fixed quark mass $m_s = 1.4 \, m_0$ with  $m_0=1.1\,$GeV.  Grey dashed line on the right shows the results for $\partial \langle L \rangle/\partial T$.}
\label{fig3}
\end{figure*}
\begin{figure*}[!ht]
\centering
\includegraphics[width=0.49\linewidth]{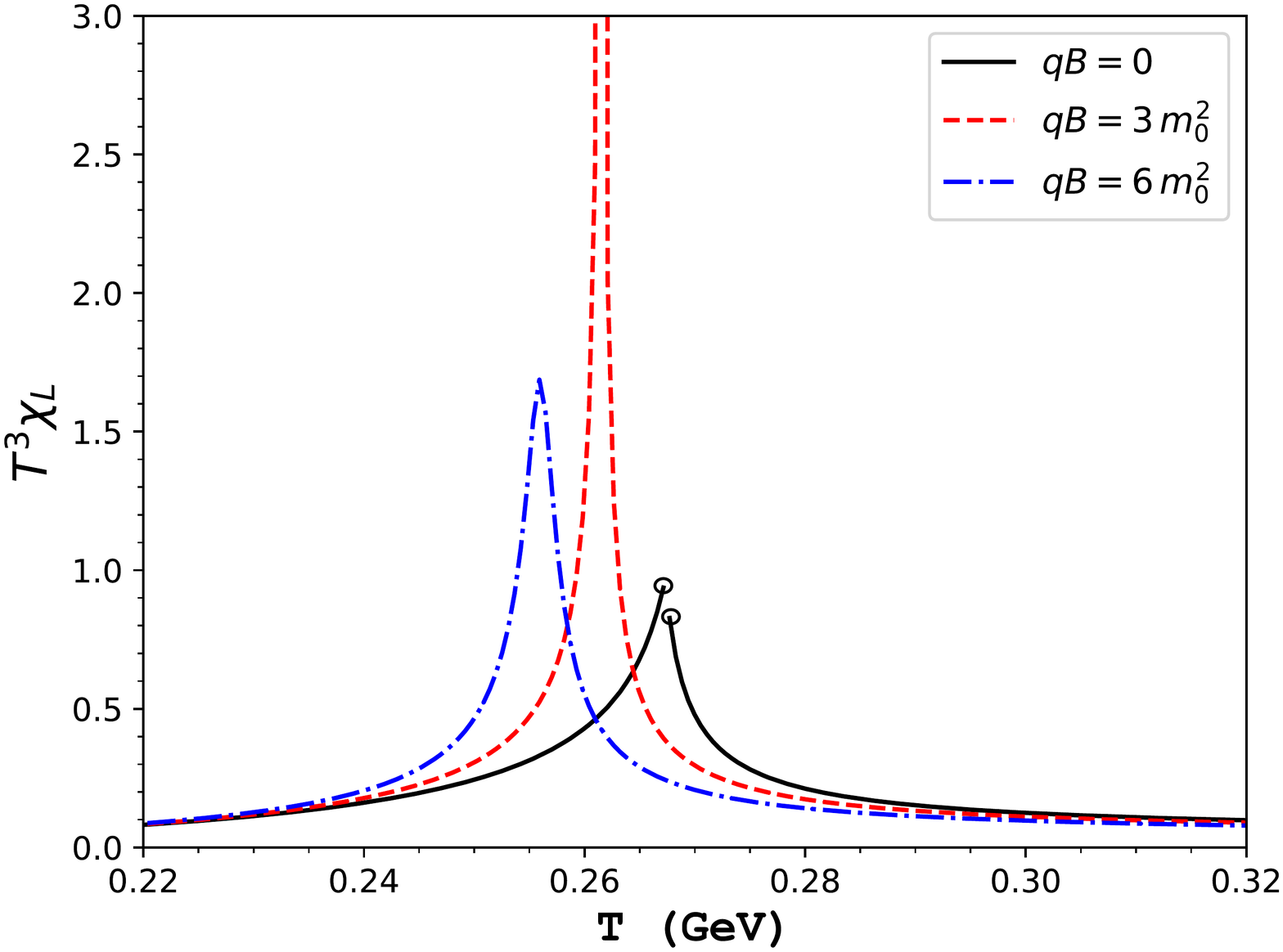}
\includegraphics[width=0.49\linewidth]{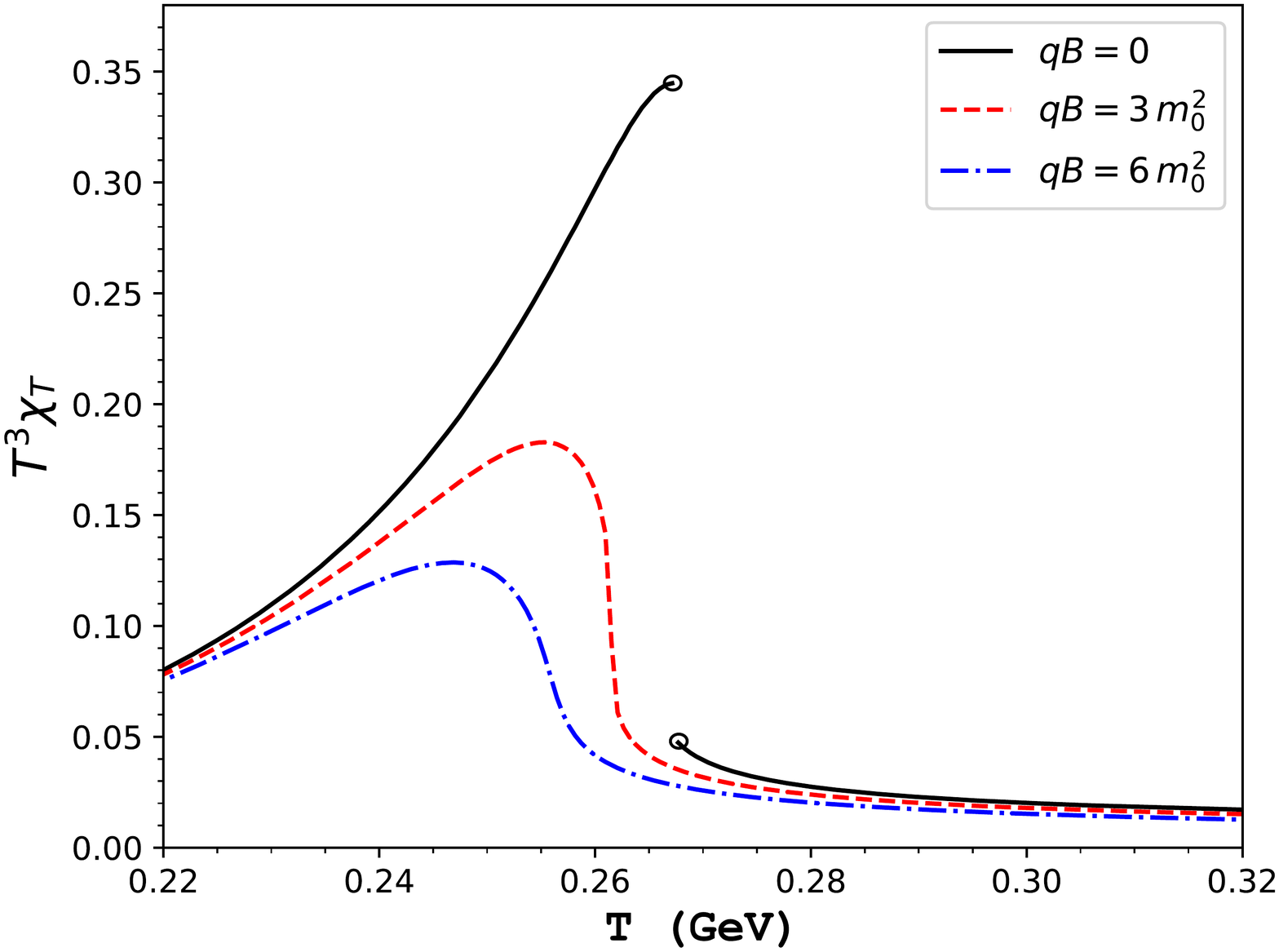}
	\caption{Similar to Fig. \ref{fig3} but for the longitudinal (left) and transverse (right) susceptibilities.}
\label{fig4}
\end{figure*}

\section{ Polyakov loop susceptibilities and their ratios }
\label{sec3}

The Polyakov loop susceptibilities measure the fluctuations of the Polyakov loops.
For Z(3) symmetry the order parameter field is complex and one can study fluctuation along the longitudinal and transverse directions.
In the language of a potential model, when one focuses on the real sector,
these susceptibilities are simply given by the inverse of the curvatures along the real and imaginary directions.
The potential $\cU_G$ in Eq.~\eqref{eq:UG} is particularly suited for the current study,
as the known susceptibilities at zero explicit breaking, i.e. the SU(3) LQCD results, are reproduced by construction.

To compute the susceptibilities in the current model,
we first solve for the expectation value of the Polyakov loop using the gap equation:

\begin{align}
  \label{eq:gap}
  \frac{\partial \cU}{\partial \phi} = 0,
 \end{align}
for $\phi = \left( x, y \right)$. The susceptibilities are then obtained from the diagonal elements of the inverse of the correlation matrix $\mathcal{C}$:

\begin{align}
  \begin{split}
	  \mathcal{C}_{ij} &= \frac{\partial^2 (U/T^4)}{\partial \phi_i \partial \phi_j}, \\
	  T^3 \, {\chi_L} &= (\mathcal{C}^{-1})_{11}, \\
	  T^3 \, {\chi_T} &= (\mathcal{C}^{-1})_{22}.
  \end{split}
 \end{align}
Note, that $\mathcal{C}$ should be evaluated using the $\phi$ that satisfies the gap equation~\eqref{eq:gap}.
The susceptibility ratio $R_T = \chi_T / \chi_L$ can be thus constructed.

The fluctuation of the absolute value of the Polyakov loop, $\chi_A$,
and the ratio $R_A = \chi_A / \chi_L$ cannot be obtained simply within a mean-field approach.
One way to compute the observable is by setting up a color group integration~\cite{Lo:2018wdo}.
Here, we provide an approximate way to determine $R_A$, based on the scaling relation discussed in Ref.~\cite{Lo:2018wdo}.

The scaling formula is based on a Gaussian approximation of the potential.
In this scheme, $R_A$ can be constructed from the mean-field results of $\ell_0, \chi_L, \chi_T$ via:

\begin{align}
	\label{eq:gauss}
  R_A(\xi, R_T) = 1 + R_T + 2 \xi^2 - \frac{2}{\pi} \, R_T \, e^{-2 \xi^2} \left[ \mathcal{F}(\xi,R_T) \right]^2,
\end{align}
with $\mathcal{F}$ given by

\begin{align}
  \begin{split}
  \mathcal{F}(\xi, R_T) =& \frac{1}{\sqrt{\pi}} \, \int_{-\infty}^{\infty} d x \, e^{-x^2 + 2 \xi x} \times  \frac{x^2}{2 R_T } \times \\
  & e^{ \frac{x^2}{2 R_T} } \times \left( K_0[\frac{x^2}{2 R_T}] + K_1[\frac{x^2}{2 R_T}] \right),
  \end{split}
\end{align}

\noindent where $K_n$ is the modified Bessel function of the second kind of the
$n$th order. The scaling variable $\xi$
is computed from~\footnote{In Ref.~\cite{Lo:2014vba}, the Gaussian fit is applied only to the pure gauge $\cU_G$. Here, an alternative scheme is used: Fitting instead the full potential $\cU$ with a Gaussian model $\cU/T^4 \approx \frac{1}{2 \, T^3\chi_L} \, (x-\ell_0)^2 + \frac{1}{2 \, T^3 \chi_T} \, y^2$}:
\begin{align}
  \begin{split}
    \xi &= \tilde{h} \times \frac{\sqrt{VT^3}}{2 \sqrt{\alpha_1}} \\
    \tilde{h} &= \frac{ { \ell_0 }}{ T^3 \, {\chi_L }} \\
    \alpha_1  &= \frac{1}{2 \, T^3 \, {\chi_L}} \\
	 R_T  &= \chi_T / \chi_L.
\end{split}
\end{align}
\noindent Note, that $R_A(T, V)$ should be computed at a finite volume to be meaningful. Otherwise $R_A \rightarrow 1$ as $V \rightarrow \infty$. In this work, we shall choose $V = (6.4 \, {\rm fm})^3$.
The numerical results of the Polyakov loop susceptibilities and their ratios will be presented in the next section.

\section{Deconfinement phase diagram in the presence of magnetic field}
\label{sec4}

\begin{figure*}[!ht]
\centering
\includegraphics[width=0.49\linewidth]{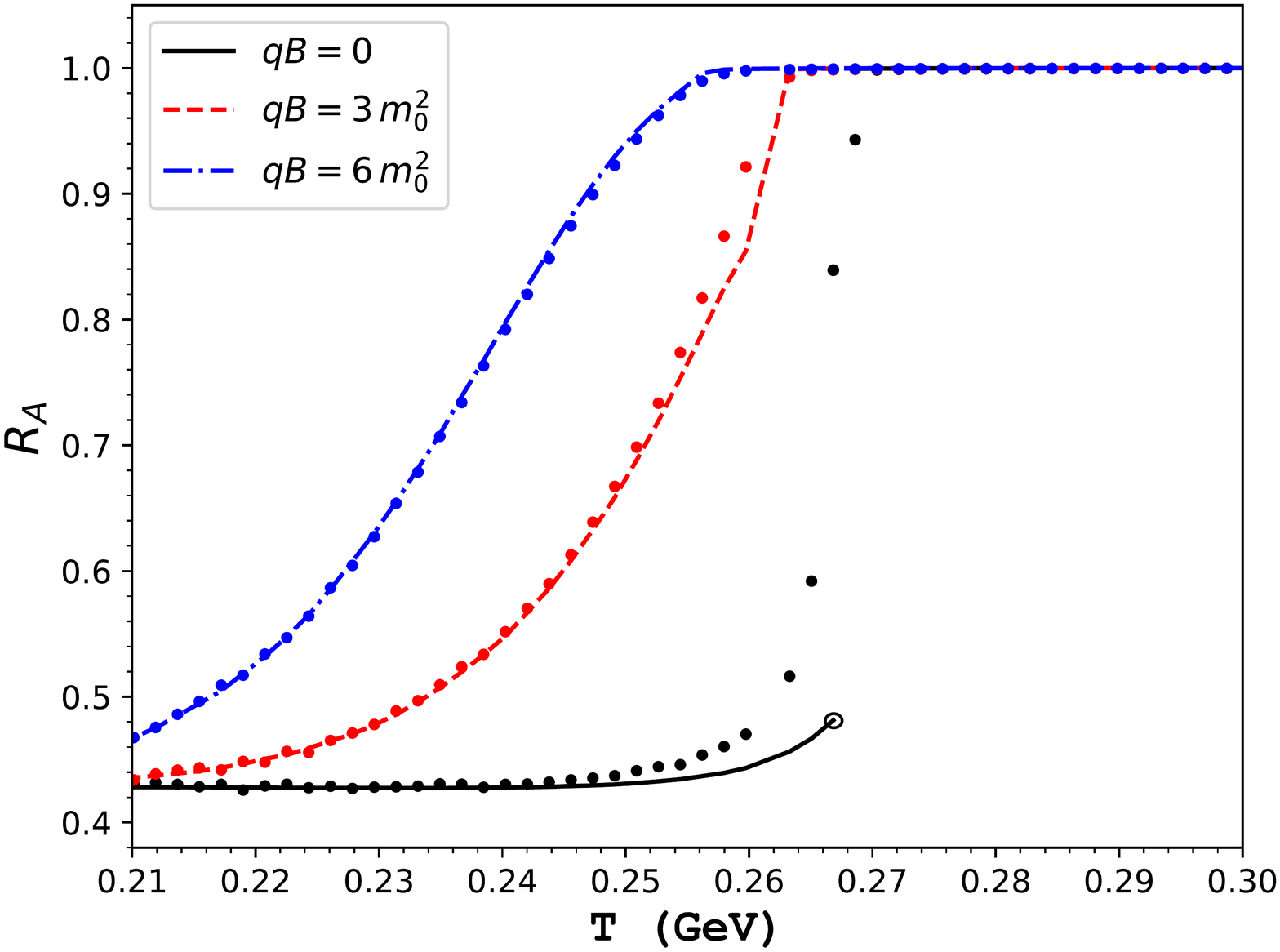}
\includegraphics[width=0.49\linewidth]{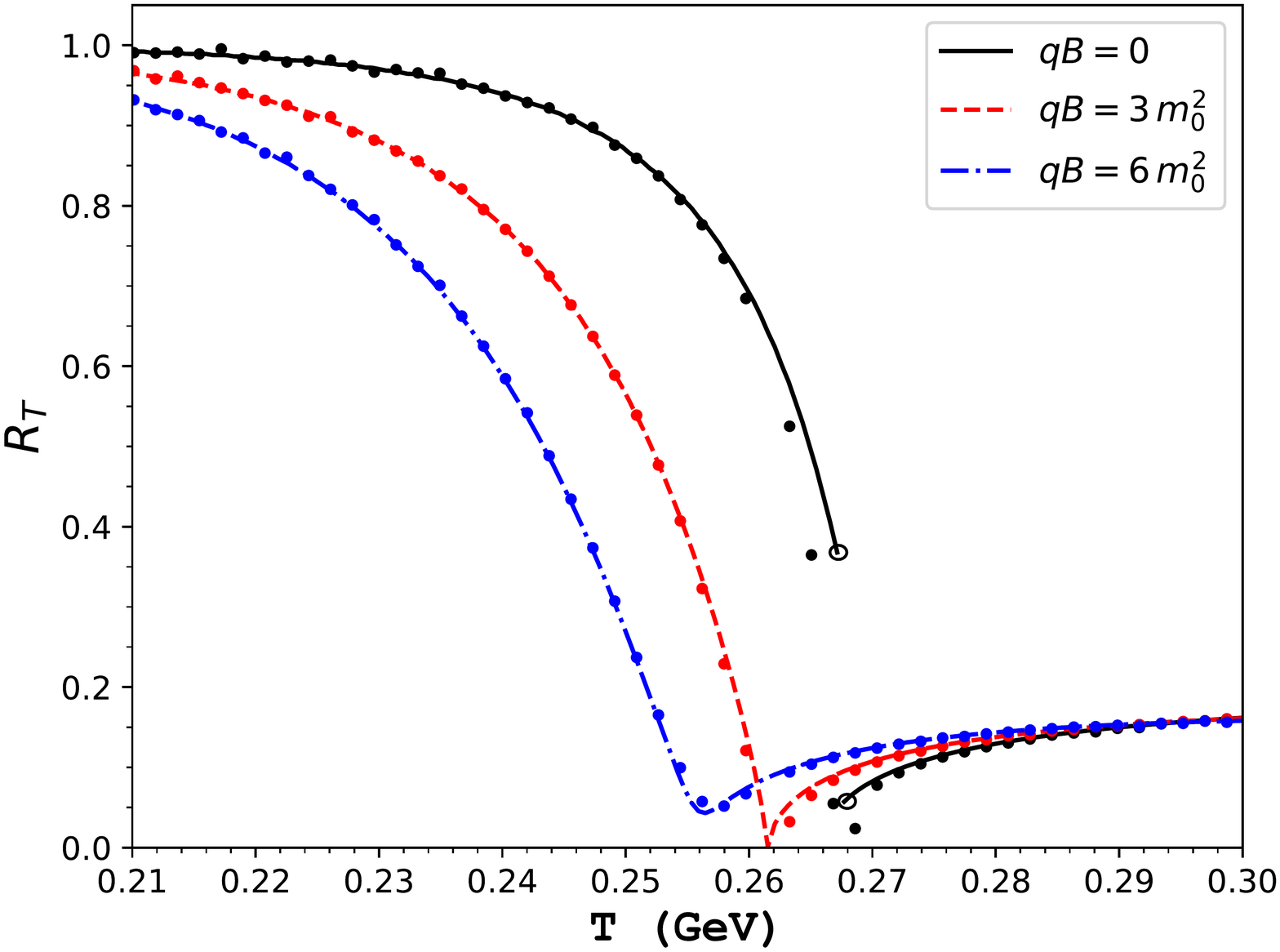}
	\caption{$R_A$ (left) and $R_T$ (right) ratios of Polyakov loop susceptibilities.
		 The points are the results obtained using a color group integration approach~\cite{Lo:2018wdo} at $V = (6.4 \, {\rm fm})^3$.
		 For $R_A$, the lines are obtained from Gaussian scaling formula in Eq. \eqref{eq:gauss}. For $R_T$, the lines are obtained from the mean-field calculations.}
\label{fig5}
\end{figure*}

\subsection{Fluctuation observables}

A key property of a first order phase transition is that it can withstand small external perturbations~\cite{Binder_1987}.
For Z(3) symmetry,  the transition remains discontinuous for sufficiently heavy quarks, and becomes continuous at the deconfinement critical point.
The critical quark mass for a single flavor system in the absence of magnetic field was found to be~\cite{Lo:2014vba}

\begin{align}
    m_0=1.1\,{\rm GeV}.
\end{align}
As shown in Sec~\ref{sec2}, the presence of an external magnetic field further enhances the explicit Z(3) symmetry breaking. This means that the critical point can be reached at a higher quark mass in the presence of $B$. To demonstrate this we consider the single flavor case (strange quark) with mass $m = 1.4 \, m_0$, for three values of magnetic field: $qB = 0$, $3 m_0^2$ and $6 m_0^2$. The observables are presented in Fig. \ref{fig3} and \ref{fig4}.

We start with the expectation value of the Polyakov loop, shown in Fig. \ref{fig3} (top left). At $B=0$ the Polyakov loop is discontinuous, indicating a first order phase transition. At a sufficiently large magnetic field $qB \approx 3 (m_0)^2$, the transition becomes continuous, and for an even larger field the transition becomes a crossover.

Similar to the LQCD studies in Refs.~\cite{Kaczmarek:2005gi,Bazavov:2016uvm}, we can calculate the entropy of a static quark $S_Q$ from the Polyakov loop within our model via

\begin{align}
	S_Q = \frac{\partial}{\partial T} \, T \ln \langle \ell \rangle.
\end{align}
The peak position of this observable was argued~\cite{Bazavov:2016uvm} to be a robust way to define the deconfinement transition temperature $T_D$.
In particular, $T_D$ thus extracted in (2+1)-QCD was found to be lower than that extracted from the inflection point of the Polyakov loop. This trend is also observed in the current model, see Fig. \ref{fig3} (top right).

In Table \ref{tab:peaks} we show the $T_D$'s extracted from the peaks of the observables: $\chi_T, \chi_L, S_Q, \frac{\partial \ell}{\partial T}$.
We also check what happens in the case of a larger Z(3) breaking: e.g. for light quarks and/or larger $B$,
and generally find the pattern: $ T_{\chi_T} < T_{S_Q} \lesssim T_{\chi_L}  < T_{\rm inflex.}$.
Differences between characteristic temperatures become substantial in this case.

The change in the order of the phase transition is also evident from the longitudinal susceptibility $T^3 \chi_L$, Fig. \ref{fig4} (left).
A key feature of this observable is that it diverges at the critical endpoint (CEP).
This makes the observable $\chi_L$ unique for defining the (pseudo) critical temperature of deconfinement.
Moreover, by studying how the magnitude of the peak changes, the critical value of $B$ can be identified for a given quark mass. On the other hand, the transverse Polyakov loop susceptibility $T^3 \chi_T$ changes smoothly across the deconfinement critical point, see Fig. \ref{fig4} (right). In this model it becomes more and more suppressed with increasing $B$.

\begin{table}[htp!]
  \begin{center}
\begin{tabular}{|c|c|c|c|c|}
  \hline
	\normalsize {}     &  $T_{\chi_T}$ (GeV)    & $ T_{\chi_L} $ (GeV) &  $T_{S_Q}$ (GeV)  &   $ T_{\rm inflex.} $ (GeV)  \\ \hline
	$qB=0$             & 0.26702                & 0.26737              & 0.26702   &   0.26737         \\ \hline
	$qB=3m_0^2$        & 0.25544                & 0.26140              & 0.26140   &   0.26175         \\ \hline
	$qB=6m_0^2$        & 0.24702                & 0.25579              & 0.25579   &   0.25614         \\ \hline
\end{tabular}
	  \caption{ Characteristic temperatures extracted from the peak positions of various observables: $\chi_T, \chi_L, S_Q, \frac{\partial \ell}{\partial T}$.
}
\label{tab:peaks}
  \end{center}
\end{table}

In Fig.~\ref{fig5} we show the ratios of the susceptibilities $R_A$ and $R_T$.
Here, we also show (as points) the results obtained from a color group integration at $V = (6.4 \, {\rm fm})^3$.
(This volume is sufficiently large to reproduce the mean-field results, e.g. $R_T$.)
The observables behave as expected.
For $R_A$ it interpolates between the two known theoretical limits from

\begin{align}
	2 - \frac{\pi}{2} \approx 0.43
\end{align}
at low temperatures to $1$ at high temperatures. The latter limit is reached more rapidly for a larger breaking strength.
We also see that the approximation formula introduced in Sec. \ref{sec3} is effective except near the transition points,
where we see substantial finite volume effects.
For $R_T$,  the Z(3) symmetry imposes limit only on the symmetric phase, i.e. $1$.
The result at high temperatures depends on the model used.
Implementing the SU(3) Haar measure is crucial to getting $R_T \ll 1$ at large temperatures, which is also the trend suggested by LQCD studies.

One theoretical motivation for studying and understanding these ratios is to seek out additional observables, other than the standard Polyakov loop, to probe deconfinement.
The issue is that the renormalized Polyakov loop extracted by LQCD is multiplied by a scheme dependent (and effectively temperature dependent) factor.
It is not clear how to match this quantity to that computed in effective models,
and calls into question the physical relevance of the deconfinement features deduced from it, e.g. the transition temperature $T_D$ extracted from its inflection point.
A similar issue also concerns a direct comparison of the susceptibilities.

\begin{figure}[!ht]
\centering
\includegraphics[width=1.\linewidth]{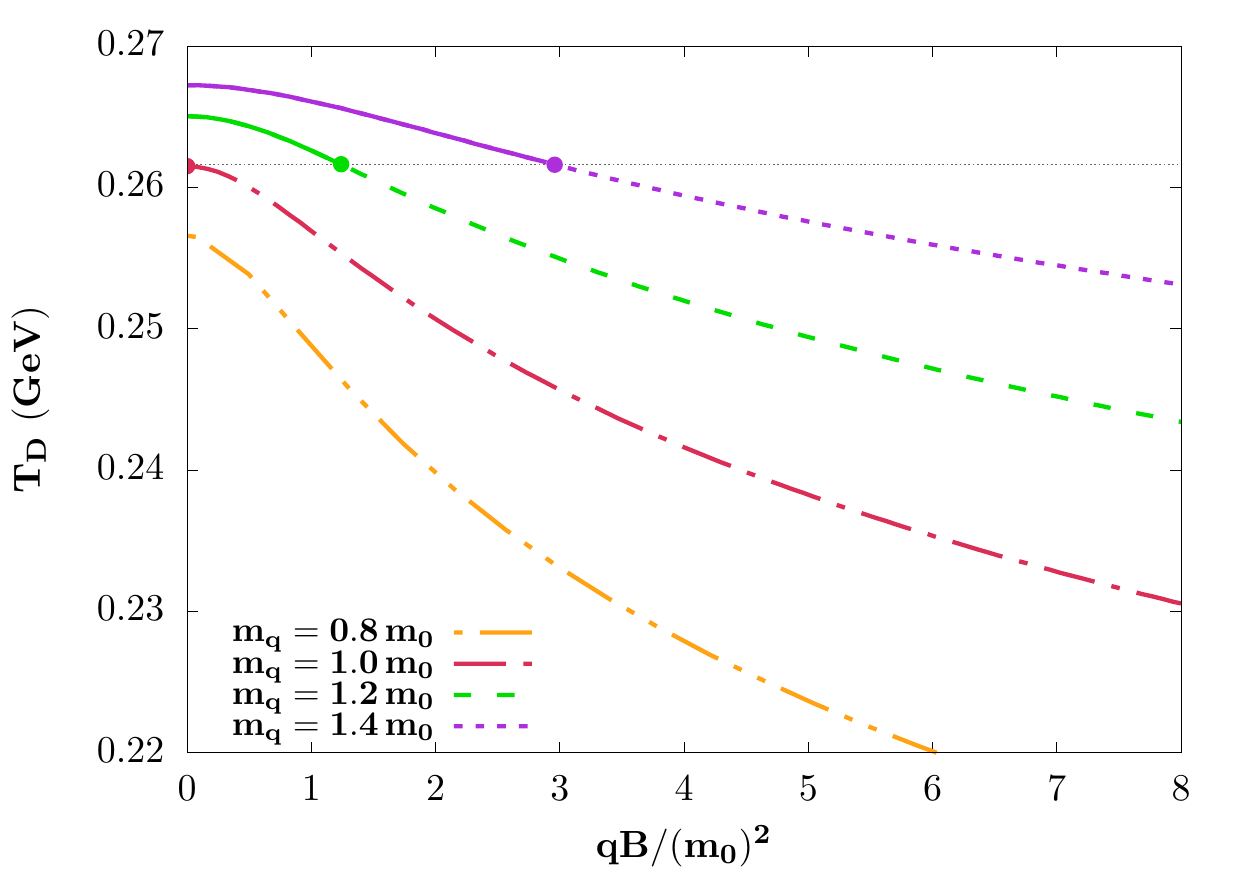}
	\caption{
 Deconfinement temperature versus  magnetic field strengths for various quark masses with  $m_0=1.10\,$ GeV. Solid lines signify first order phase transitions, dashed lines correspond to crossover, and dots denote  the deconfinement CEP's
at $T_{CEP} \approx 0.261$ GeV.
}
\label{fig6}
\end{figure}

\begin{figure*}[!ht]
\centering
\includegraphics[width=0.49\linewidth]{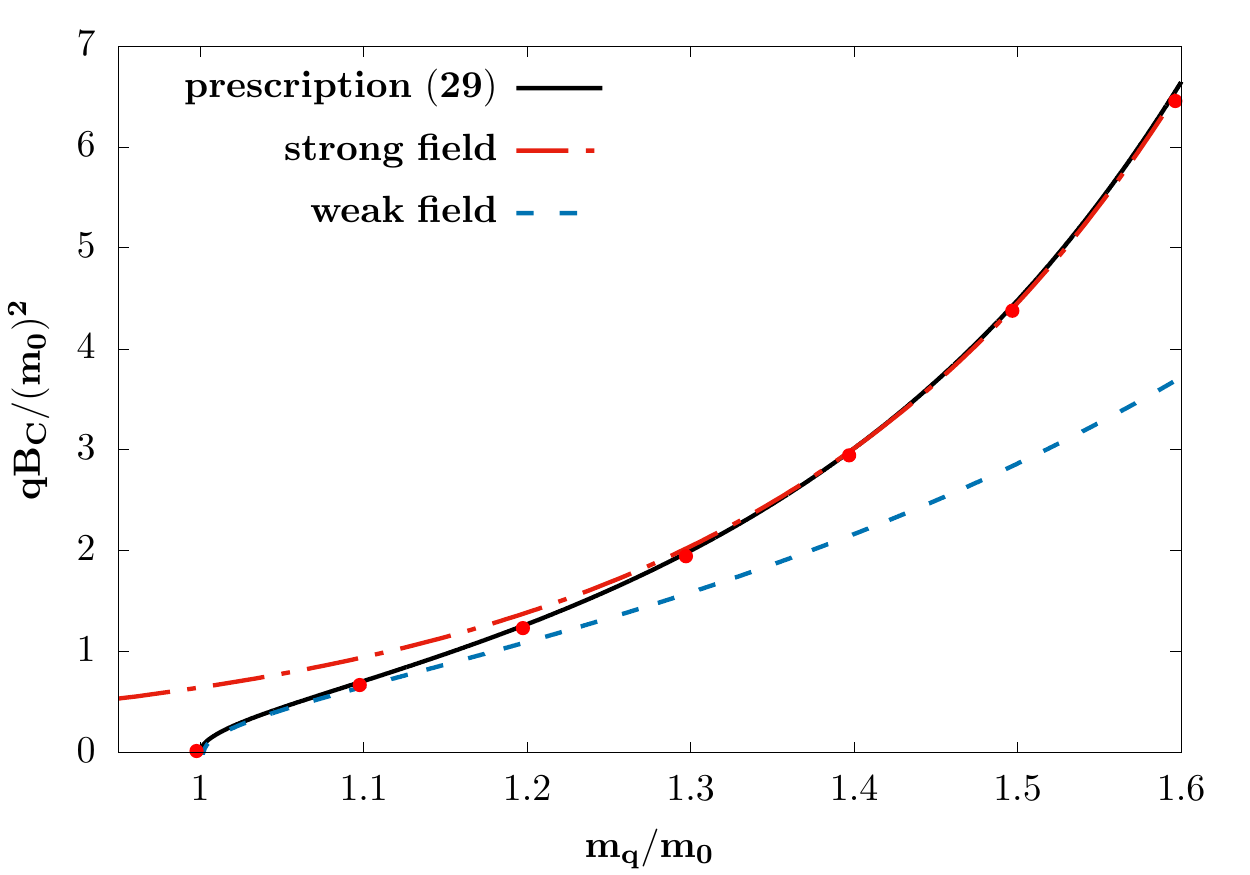}
\includegraphics[width=0.49\linewidth]{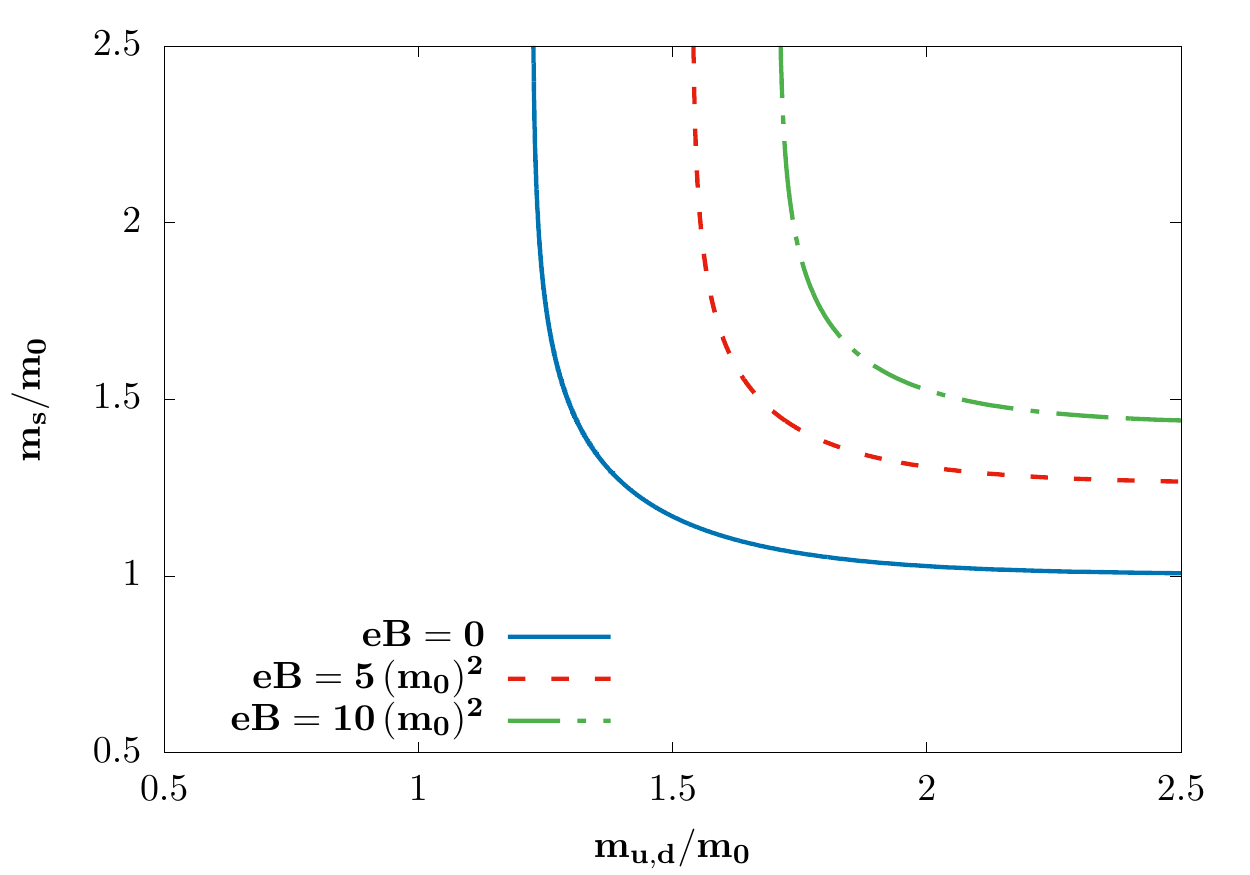}
	\caption{
Left: lines of critical points (Eq. \eqref{eq:crit}), represented by the critical $B$ versus quark  masses
 for a single flavor system. Red points signify  the location of  diverging peak positions   of $\chi_L$. The blue-dashed and
red-dash-dotted lines  are obtained by using the weak and strong field limits of $h^B_Q$ in Eq.~\eqref{eq:key1}, respectively.
Right:
 the dependence of the phase structure  on the different  quark  masses  and magnetic field strengths in (2+1)-flavor system   with physical electric charges. Lines indicate the locations of CEP's.
}
\label{fig7}
\end{figure*}

One possibility is to consider dual parameters, e.g. the dressed Polyakov loop~\cite{Bilgici:2008qy}, which is 
calculated from the chiral condensate with a varying boundary condition of quarks. 
Note that the temperature derivatives of these quantities can be quite different from those of the Polyakov loop~\cite{Morita:2011jv}. 
Another alternative is to study ratios of susceptibilities.
The idea is that by constructing ratios the multiplicative renormalization factor drops out \cite{Lo:2013etb}.
However, the task of extracting useful information from these quantities turns out to be more involved than originally thought \cite{Lo:2013hla}.
In particular, the proper renormalization of the susceptibilities may not be only multiplicative, if true even the ratios would not be scheme independent.
As the magnetic field provide an additional handle to study these ratios, the model predictions presented here may
guide future LQCD studies at finite magnetic field.

\subsection{Phase diagram}

To further investigate the effect of a magnetic field on deconfinement we calculated $T_D$ by tracking the peak of $\chi_L$ at various magnetic field strengths.
The results are shown in  Fig.~\ref{fig6} for the single flavor system, and with quark masses,  $m_s=(0.8, 1, 1.2, 1.4) \, m_0$.

The deconfinement temperature $T_D$ in this model generally decreases when $h_Q^B$ increases. Thus, when plotted as a function of $B$,
the heavier quarks would require a larger magnetic field strength compared to the light ones, to reach the same $T_D$.
This simple observation explains the main features seen in Fig.~\ref{fig6}.

Following the arguments in Ref.~\cite{Lo:2014vba}, for effective models of this class,
the temperature corresponding to the  deconfinement critical endpoint $T_{CEP}$ would remain constant
\footnote{We briefly recount the arguments. For potential of the form $\cU = \cU_G - h x$,
the condition for CEPs reads:
\begin{align}
        \frac{\partial {\cU}_G}{\partial x} = h, ~~
        \frac{\partial^2 {\cU}_G}{\partial x} = 0, ~~{\rm and} ~~
        \frac{\partial^3 {\cU}_G}{\partial x} = 0.
\end{align}
The solution fixes the critical values of the Polyakov loop $x_{CEP}$, $h_c$, and $T_{CEP}$.
If ${\cU}_G$ is independent of the external fields, the latter two conditions uniquely determine $T_{CEP}$ and $x_{CEP}$,
making $T_{CEP}$ insensitive to the external fields.}
\begin{align}
T_{CEP}\approx 0.261\,\text{GeV}.
\label{eq:T_CEP}
\end{align}
This is verified for the current case of a finite magnetic field, as the
 $B$-dependence enters only through the explicit breaking term.

Under the same assumptions, the position of deconfinement critical point can be determined from the following condition:
\begin{align}
 \sum\limits_{f=u,d,s}h^B_Q(m_f,T_{CEP},q_f B)=h_c,
 \label{eq:crit}
 \end{align}
where $T_{CEP}$ is given in Eq. \eqref{eq:T_CEP} and
\begin{align}
h_c\approx0.144,
\end{align}
is the critical breaking strength.

We demonstrate two uses of Eq. \eqref{eq:crit}.
First, this dictates the critical strength of the magnetic field required to reach the deconfinement CEP for quark masses above $m_0$. This is shown by the solid black line on Fig. \ref{fig7} (left) for the single flavor system,
bounded by two limiting lines obtained via the weak and strong field approximations of $h^B_Q$ in Eq.~\eqref{eq:key1}.
This agrees with the results obtained by tracking the diverging peaks of $\chi_L$, shown as red points in the same figure.

Second, when $m_s$ and $m_l$ are considered as independent variables, condition \eqref{eq:crit} determines a critical surface in the $(m_l,m_s)$ plane~\cite{Kashiwa:2012wa, Kashiwa:2013rm}.
See also Refs.~\cite{whot, Fromm:2011qi, Ejiri:2019csa} for the LQCD determination of this graph. (The continuum extrapolation, however, remains elusive.)
For a rough comparison to a recent work~\cite{Ejiri:2019csa}, the study reported $m_{PS}/T_{CEP} = (15.73, 11.15)$ for $N_t = (4, 6)$ in 2-flavor QCD system (NLO).
In our 2-flavor model, taking the pseudoscalar mass $m_{PS} \approx 2 m_{\rm crit.}$, with $m_{\rm crit.} \approx 1.353$ GeV and $T_{CEP} \approx 0.261$ GeV,
the ratio $m_{PS}/T_{CEP}\approx 10.37$. 
The matrix model~\cite{Kashiwa:2013rm} gives a comparable value of $11.8$. 
This encourages a closer comparative study of effective models and LQCD calculations.

The effect of an external magnetic field on the phase diagram is shown on Fig. \ref{fig7} (right).
Similar to a quark chemical potential~\cite{Lo:2014vba}, an increasing magnetic field tends to shrink the region of the first order phase transition.

\section{Toward including dynamical light quarks}
\label{sec5}

It is known that a naive implementation of the PNJL model would give a $T_D$ that increases with $B$.
This fact can be easily understood in the current model.

In previous sections, we have shown,  that
$h_Q^B(m, T, B)$
is an increasing function of $B$ at fixed $(m, T)$. In addition, it decreases as $m$ increases.
In the PNJL model, the quark mass $m$ is to be substituted with the constituent quark mass $M_Q(T, B)$.
This brings in,  an additional magnetic field dependence, and $h_Q^B$ should be viewed as a functional of $M_Q(T,B)$:

\begin{align}
h_Q^B[M_Q(T, B), T, B].
\end{align}

Most NJL models can capture the effect of magnetic catalysis, i.e. $M_Q(T \approx 0, B)$ increases with $B$.
This supersedes the enhancing effect of the explicit $B$ dependence on $h_Q^B$ and leads to a smaller explicit Z(3) breaking in a broad temperature range, including the vicinity of $T_D$.
We thus expect the trend of an increasing $T_D$ with $B$ for this class of models.

\begin{figure*}[!ht]
\centering
\includegraphics[width=0.49\linewidth]{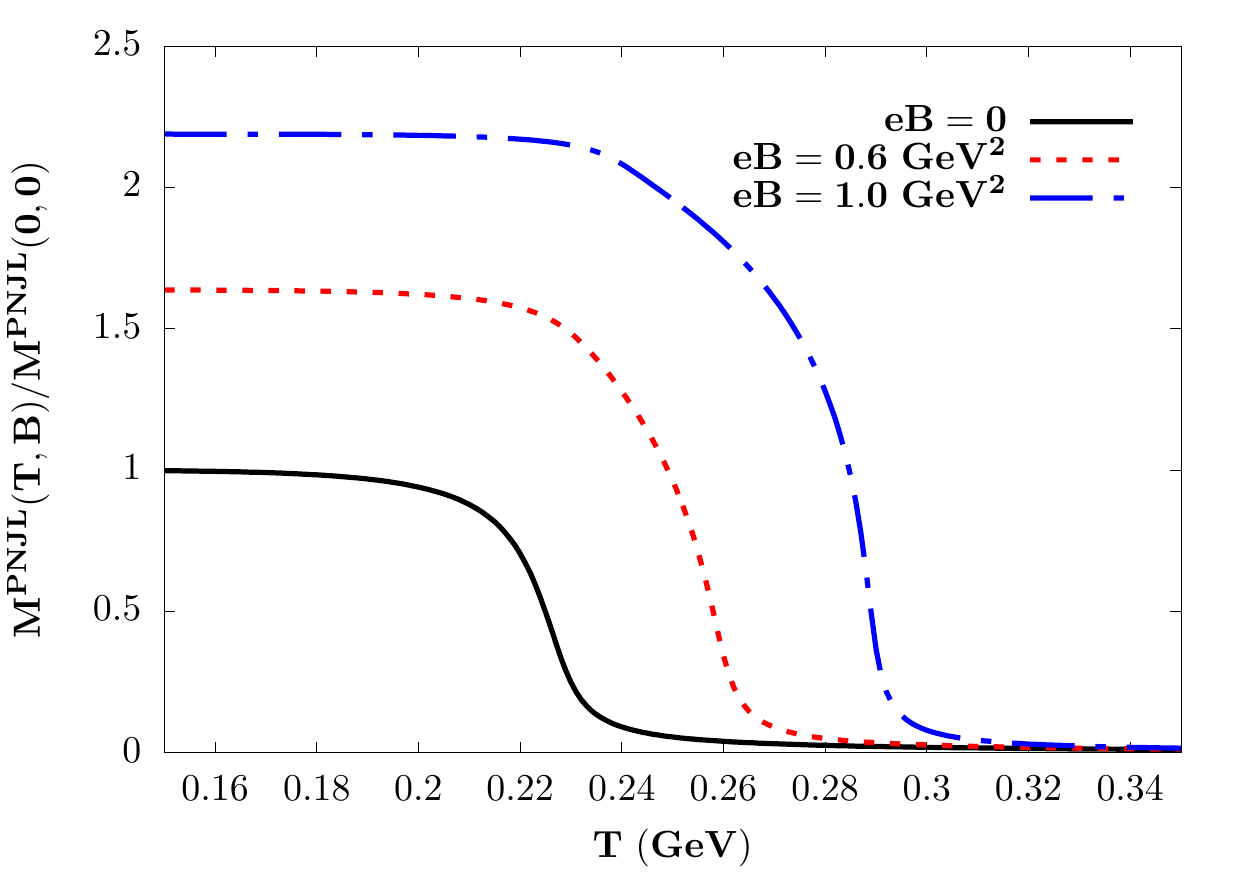}
\includegraphics[width=0.49\linewidth]{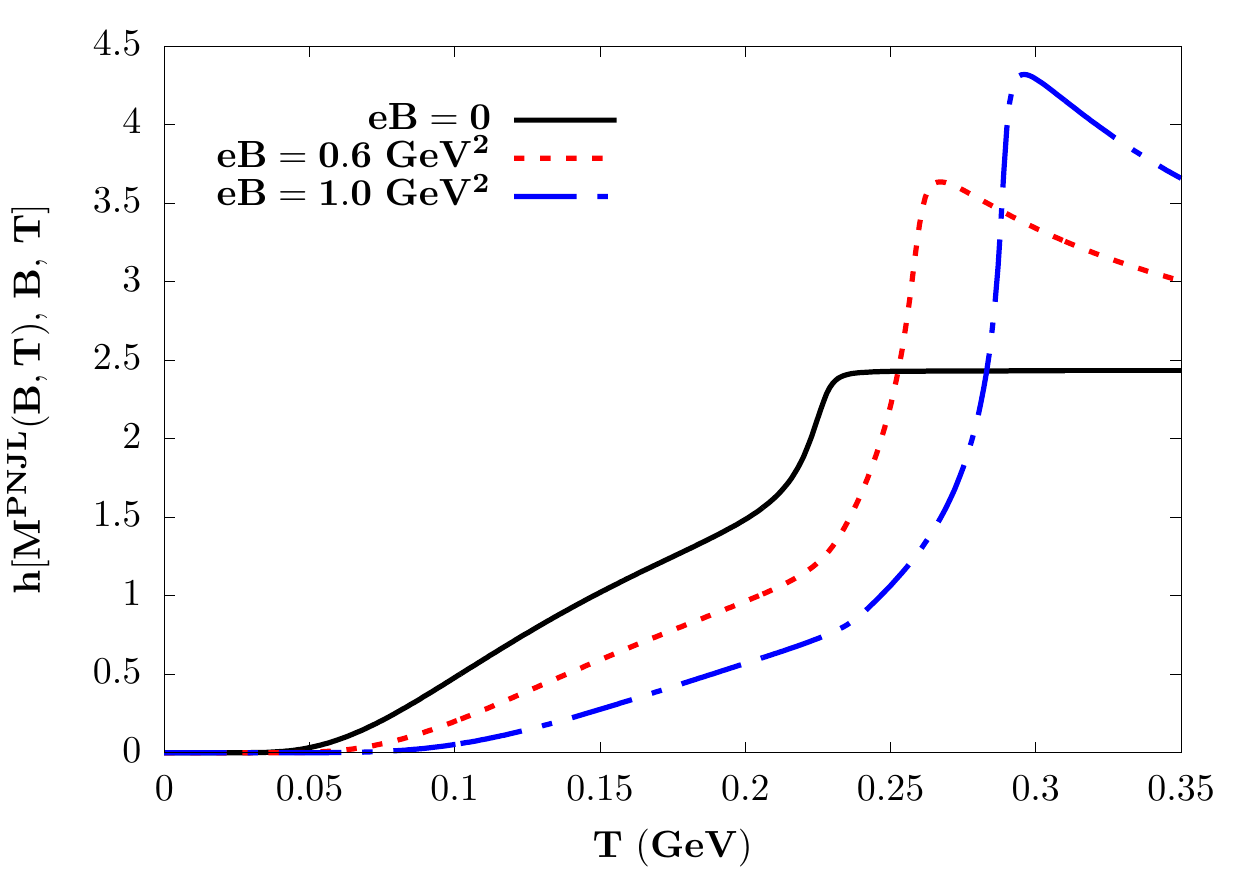}
	\caption{Constituent quark mass (left) and the corresponding Z(3) breaking $h_Q^B[M(T,B), T, B]$ (right) in  the  PNJL model, at various  strengths of   magnetic fields $B$. The NJL model follows from Ref.~\cite{Boomsma:2009yk} and the pure glue potential $\cU_G$ from Eq.~\eqref{eq:UG}.}
\label{fig8}
\end{figure*}

We demonstrate this fact by performing a (2-flavor) PNJL model calculations, based on the NJL model in Ref.~\cite{Boomsma:2009yk}~\footnote{
	We fine-tuned the NJL model parameters to reproduce the LQCD results~\cite{Bali:2012zg} at $T=0$, finite $B$. See Appendix~\ref{app1}.
In the text, we perform calculations in the minimal ($u,d$)-quarks splitting case of Ref.~\cite{Boomsma:2009yk}. 
The maximal splitting case is discussed in the Appendix~\ref{app1}.}
 and the pure glue potential $\cU_G$  in Eq.~\eqref{eq:UG}.
The constituent quark mass functions at various values of $B$, and the Z(3) breaking strength $h_Q^B$ computed from them, are shown in Fig.~\ref{fig8}. As $B$ increases, the diminishing effect of the larger quark mass on $h_Q^B$ overrides the enhancing effect from the explicit $B$ dependence, suggesting a lower Z(3) breaking.
The trend of an increasing $T_D$ with $B$ is realized even when the full potential $\cU_Q$ is employed (as in this self-consistent PNJL model calculation), as seen in Fig.~\ref{fig10}.

\begin{figure*}[!ht]
\centering
\includegraphics[width=0.49\linewidth]{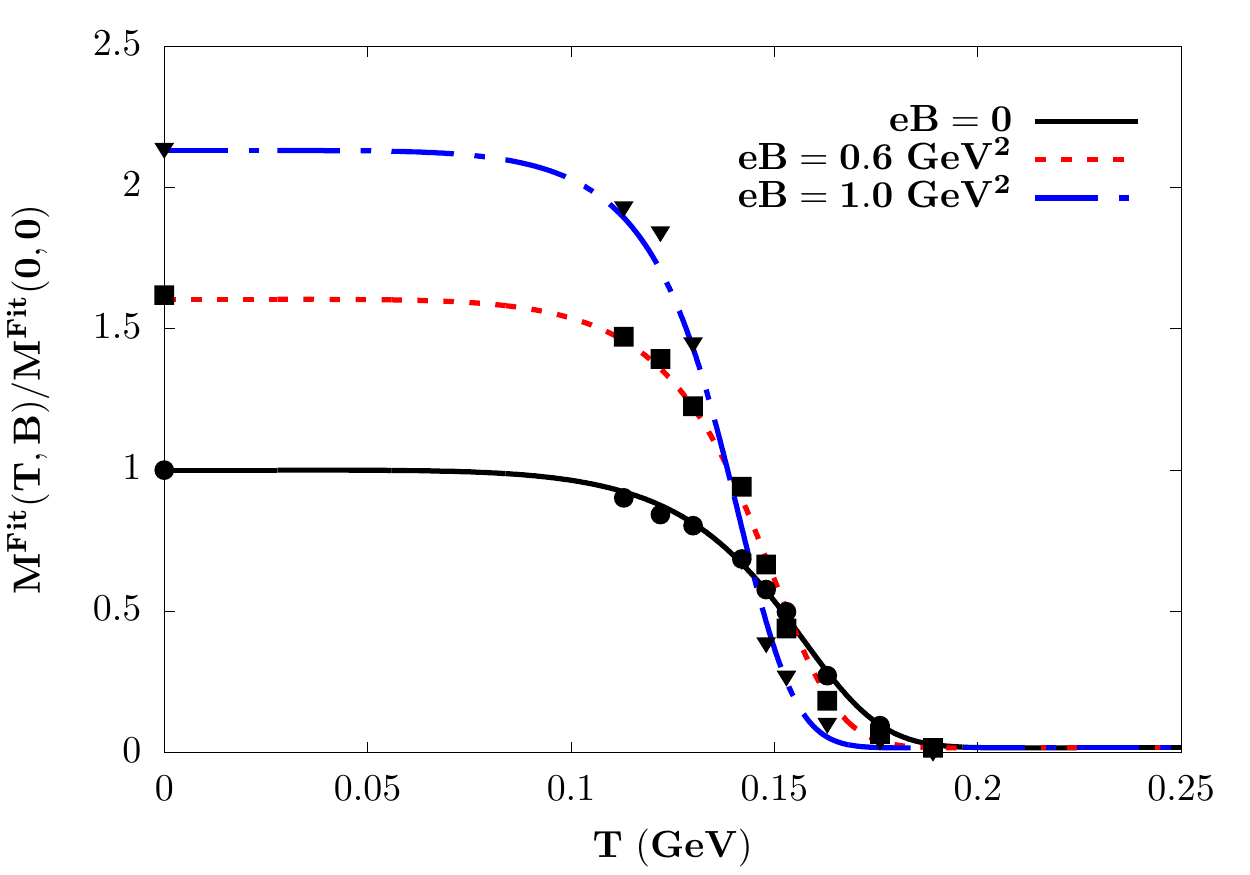}
\includegraphics[width=0.49\linewidth]{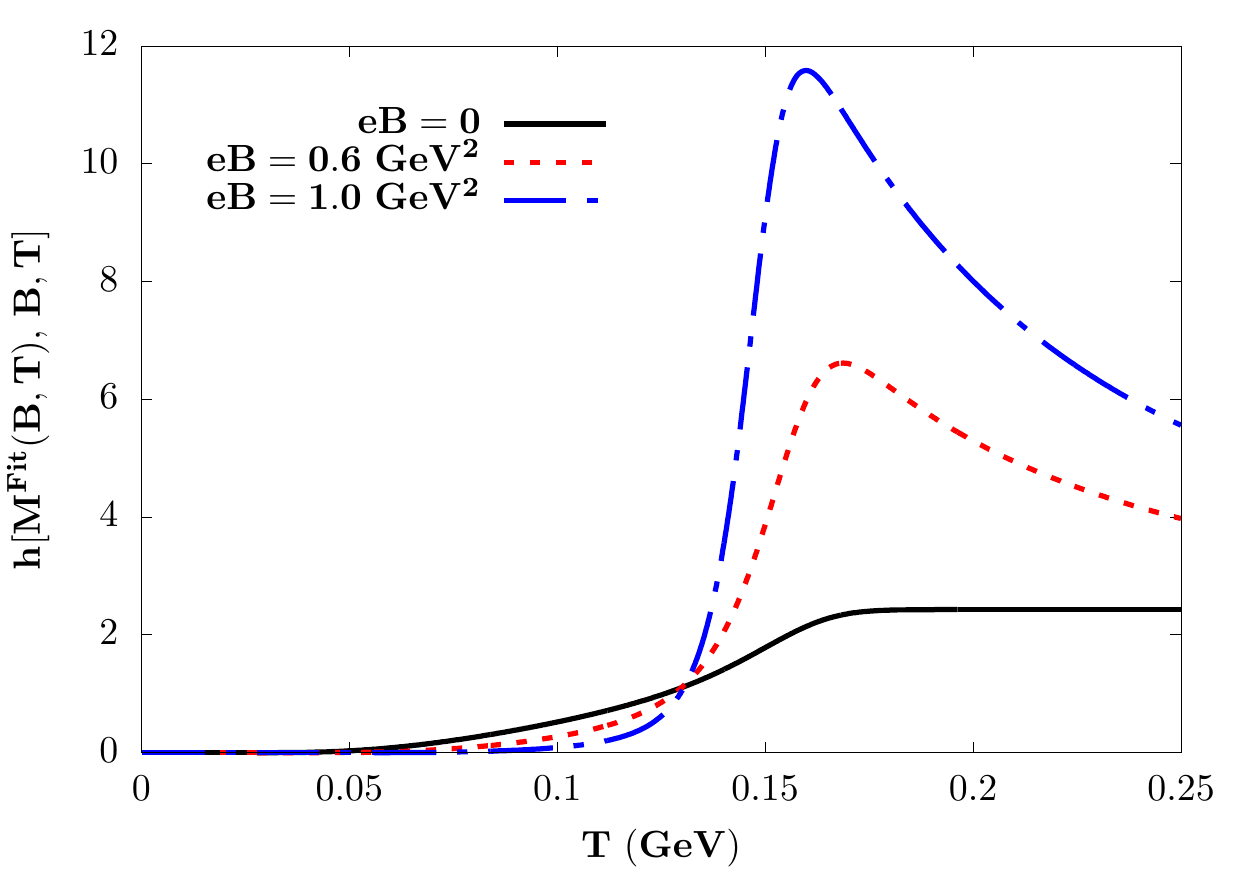}
	\caption{Similar to Fig.~\ref{fig8} but with the constituent quark mass function from  Eq.~\eqref{eq:latpnjl} based on the  fit to  LQCD data on chiral condensate~\cite{Bali:2012zg}.  Data points in  the left-hand figure  are computed based on Eq.~\eqref{eq:latpnjl} and the LQCD results (function $\mathcal{F}(T,B)$) from Ref.~\cite{Bali:2012zg}.}
\label{fig9}
\end{figure*}

\begin{figure}[!ht]
\centering
\includegraphics[width=1\linewidth]{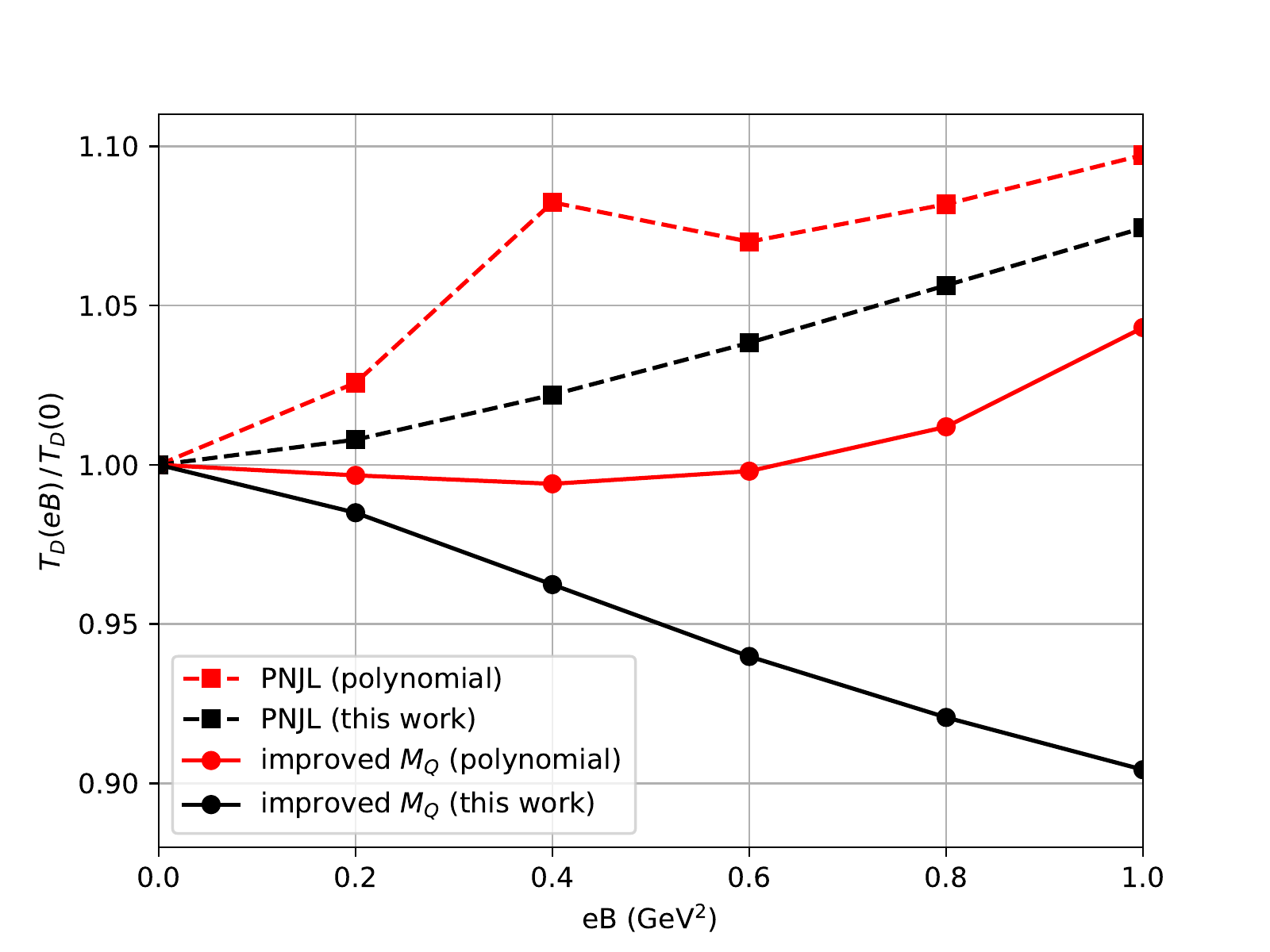}
	\caption{Deconfinement temperature $T_D$ (normalized to the value at vanishing $B$ in the corresponding model) versus  magnetic field $B$ for quark mass functions based on the  PNJL  and the  LQCD-improved models.
	         The difference in quark mass functions implies  essential differences in the Z(3) breaking strength $h_Q^B[M_Q(T,B), T, B]$,
		 which  consequently result in  the opposite  trends in  $T_D$ dependence on $B$.}

\label{fig10}
\end{figure}

We now explore, under the same mechanism, whether an improved $M_Q(T, B)$ would give the correct trend of a decreasing $T_D$ with $B$, as observed in LQCD calculations, e.g. Ref.~\cite{Endrodi:2019zrl}. To this end, we employ the following parametrization of $M_Q(T,B)$:

\begin{align}
	\label{eq:latpnjl}
	M_Q(T, B) = m - 2 G \, \langle \bar{\psi} \psi \rangle_0 \mathcal{F}(T, B)
\end{align}
where $m = 6$ MeV and $G \Lambda^2 = 2.435$, $\Lambda = 0.51507$ GeV, giving $\langle \bar{\psi} \psi \rangle_0 = - 2 \times (211 \, {\rm MeV})^3$.

The essential difference here is,  that we employ the lattice result~\cite{Bali:2012zg} on the ratio of chiral condensates to obtain $\mathcal{F}$:

\begin{align}
	\mathcal{F}(T, B) = \langle \bar{\psi} \psi \rangle(T, B)/\langle \bar{\psi} \psi \rangle_0.
\end{align}

Based on the LQCD results on the quark condensate at zero temperature~\cite{Bali:2012zg}, i.e. $\mathcal{F}_0(B) = \mathcal{F}(T=0,B)$,
and the chiral transition temperature at finite $B$, i.e. $T_\chi(B)$~\cite{Bali:2011qj},
we construct a robust parametrization of the function $\mathcal{F}(T, B)$:

\begin{align}
	\label{eq:bigparam}
	\begin{split}
		\mathcal{F}(T, B) &= \frac{\mathcal{F}_0(B)}{\mathcal{F}_1(T, B)} \\
		\mathcal{F}_0(B) &= 1 + \frac{1}{2} \, \sum_{f=u,d} \, a_1 \, (\sqrt{1+a_2(q_f B)^2}-1) \\
		\mathcal{F}_1(T,B) &= \frac{\alpha(B) + e^{ 2 \, (T/T_\chi(B))^6 }}{1 + \alpha(B)},
	\end{split}
\end{align}
with

\begin{align}
	\begin{split}
		a_1 &= 0.257  \\
		a_2 &= 115.5 \\
		\alpha(B) &= 2.47 + 4 (eB)^2 \\
		T_\chi(B) &= 0.159 - \frac{0.0326 \, (eB)^2}{1 + 0.4 \, (eB)^6},
	\end{split}
\end{align}
where all quantities are in appropriate units of ${\rm GeVs}$. 
The parametrization is restricted to $eB \lesssim 1 \, {\rm GeV }^2$. 
Note that at $T=0$, $\mathcal{F}(T=0, B) \rightarrow \mathcal{F}_0(B)$.
The shape function $\mathcal{F}_0(B)$ describes the increase of chiral condensate with $B$ at $T=0$. 
The functional form is derived from an NJL model. Details can be found in Appendix~\ref{app2}.

The resulting $M_Q(T, B)$'s are shown in Fig.~\ref{fig9}.
At low temperatures, the mass functions exhibit a similar increase with $B$ as in the PNJL model case.
The key feature, however, is a faster drop of $M_Q$ with temperature.
This restricts the diminishing effect of a large quark mass, and instead, the enhancing effect from the explicit $B$-dependence takes over.
A direct calculation shows, that indeed $h_Q^B$ is strengthened   by $B$ in the essential temperature range.
Employing such $h_Q^B$ as the explicit Z(3) breaking potential,  gives the trend of a decreasing $T_D$ with $B$, see Fig.~\ref{fig10}.
~\footnote{
Unfortunately,  using the full $\cU_Q$ potential
would reduce the explicit Z(3) breaking compared to the linear term, as shown in Fig.~\ref{fig1} (right),
counteracting the effect of an improved $M_Q$.
This suggests additional modifications of the Polyakov loop potential beyond the one-loop $U_Q$ are required
to effectively enhance the Z(3) breaking.
The same observation is made already for the $B=0$ case~\cite{Lo:2018wdo}.
}

To examine the dependence on the Polyakov loop potential,
we also perform the above analysis for the polynomial potential in Ref.~\cite{plm1}.
The result on $T_D$ versus $B$ is shown in Fig.~\ref{fig10}.
When used in a PNJL model, the polynomial potential gives a substantially stronger rising trend in $T_D(B)/T_D(0)$ as $\cU_G$ in Eq.~\eqref{eq:UG}.
When the improved quark mass function $M_Q(T,B)$~\eqref{eq:latpnjl} is used, the corresponding $T_D(B)/T_D(0)$ shows the correct decreasing trend for $eB < 0.5$ GeV, though substantially weaker, and eventually rises again.~\footnote{A similar behavior of $T_D(B)/T_D(0)$ was also reported when a $B$-dependent coupling is employed~\cite{Fraga:2013ova}.} This clearly demonstrates the merit of using an improved Polyakov loop potential, where the locations and the curvatures around minima are properly adjusted.

With this general argument via $h_Q^B[M_Q(T,B), T, B]$,
we have demonstrated the delicate interplay between chiral dynamics (for the correct $M_Q(T,B)$) and deconfinement.
This could help in constraining the missing interactions in effective chiral models.
Moreover, a further study could investigate how the explicit and implicit (via $M_Q(T,B)$) $B$-dependences of $h_Q^B$ (and the higher order terms)
are related to the effects from valence and sea quarks~\cite{Bruckmann:2013oba}.
While the valence quarks always enhance the chiral condensate as $B$ increases,
the sea quarks are found to reduce it in temperatures near the chiral transition.
Their effects on Z(3) symmetry breaking is left for future research.

\section{Conclusions}
\label{sec6}

It has been demonstrated that an external magnetic field tends to strengthen the explicit Z(3) breaking.
This is a general feature of the one-loop fermionic determinant term.
A schematic mean-field calculation shows that the deconfinement phase transition is enhanced, as seen from the lowering of the critical temperature at fixed $m$ or
the increase in the critical quark mass.
A compact phenomenological formula is derived to capture the effects of a finite magnetic field $B$ on Z(3) breaking:
(i) At small $B$, the correction term $\propto B^2$;
(ii) At large $B$, the lowest Landau level dominates and the breaking strength $h \propto B$.
Extrapolating the results to the case of zero quark masses, we find also
 terms like $B^2 \ln B$ and $B^3$.

In addition to quark masses, the magnetic field provides another handle to study Z(3) symmetry breaking.
While we demonstrated the inclusion of the higher order terms of the one-loop fermionic potential within mean-field treatments does not lead to substantial changes in the Z(3) breaking strength, we do expect additional modifications from the changes in the Polyakov loop potential, and various beyond mean-field effects, e.g. the spatial dependence of the Polyakov loops and their correlators. We hope that a more detailed study of the fluctuation observables within LQCD can help to distinguish the various effects.

From the perspective of continuum models, the general effects of a finite magnetic field on Z(3) breaking is similar to that of a finite chemical potential.
However, for LQCD studies it makes a huge difference: the former would present no sign problem.
In particular, we propose the study of the following observables. First,
\begin{align}\label{pok}
	\frac{\partial}{\partial B} \, \ln [\langle \ell \rangle(m, T, B)],
\end{align}
since the logarithm removes the multiplicative renormalization factor from the extracted Polyakov loop.
In a continuum model,
\begin{align}
	\frac{\partial}{\partial B} \, \ln [\langle \ell \rangle(m, T, B)] = \frac{T^3 \, \chi_L}{\langle \ell \rangle} \times \frac{\partial h_Q(m, T, B)}{\partial B}.
\end{align}
The second suggestion is to study the magnetic field dependence of the susceptibilities and their ratios, as worked out in this paper.
Note,  that from the color group integration approach~\cite{Lo:2018wdo} we expect $\chi_A$ and $R_A$ to be explicitly volume dependent.
To extract useful information from these quantities, they need to be studied either at a finite volume, or as a function of the scaling variable $\xi$.

Finally,  calculations 
 of the correlation function~\cite{Bonati:2017uvz,Rucci:2019hcd} (versus distances) and establishing the relation between correlation functions and the susceptibilities will guide the study of
the spatial- or momentum-dependence of the Polyakov loop fields.
The latter is being explored within the current potential model and will be reported elsewhere.

\begin{acknowledgments}
 We acknowledge the support by the Polish National Science Center (NCN) under the Opus grant no. 2018/31/B/ST2/01663. K.R. also  acknowledges partial support  from   the  Polish  Ministry  of  Science  and  Higher Education and  stimulating  discussions with Bengt Friman, Frithjof Karsch and  Swagato Mukherjee.
 \end{acknowledgments}

\appendix

\section{Explicit Z(3) breaking strength in the limit of weak and strong magnetic fields}
\label{app1}
In this appendix, we derive the expressions for the explicit Z(3) breaking field in the limits of strong
and weak magnetic field, presented in Eq.~\eqref{eq:key1} and Eq.~\eqref{eq:key2}.

To facilitate the discussion, we introduce variables $x = m/T$ and $dy = 2 \vert qB \vert / T^2$.
The full expression of $h_Q(x, dy)$ is given by
\begin{align}
	\label{eq:sum}
	h_Q^B(x,dy)=\frac{3}{2 \pi^2} \, dy \, \sum\limits_{\sigma=0,1}\sum\limits_{k=0}^\infty \bar{m}_B(k, \sigma) \, K_1\left( \bar{m}_B(k, \sigma) \right)\,,
\end{align}
where
\begin{align}
	\bar{m}_B(k,\sigma)= \sqrt{x^2 + (k+\sigma) \, dy}.
\end{align}
This expression can be readily computed numerically for arbitrary values of $(x, dy)$.
Nevertheless,  the strong and weak field limits provide an intuitive way to understand its behavior.
\begin{figure*}[t]
\centering
\includegraphics[width=0.49\linewidth]{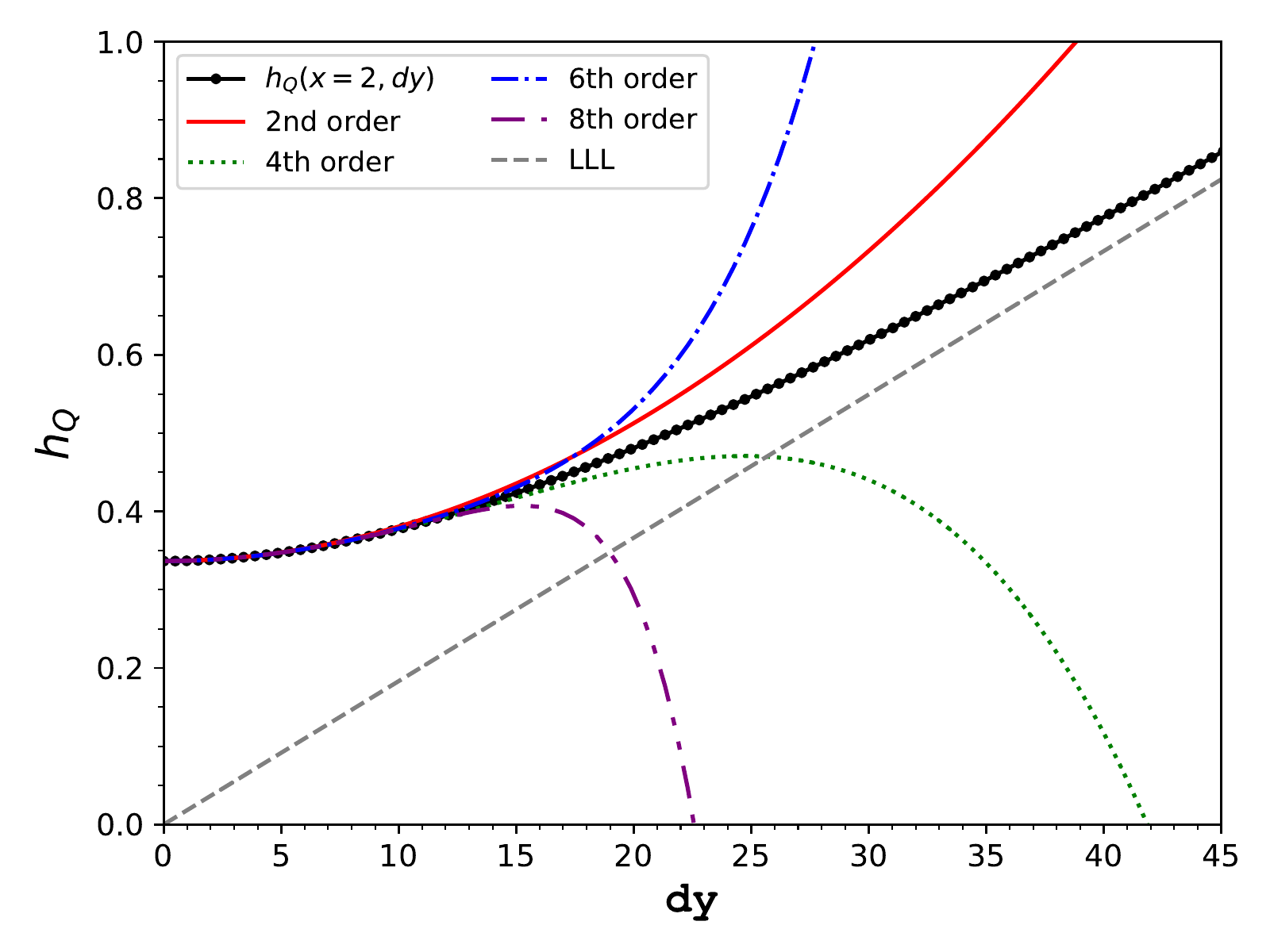}
\includegraphics[width=0.49\linewidth]{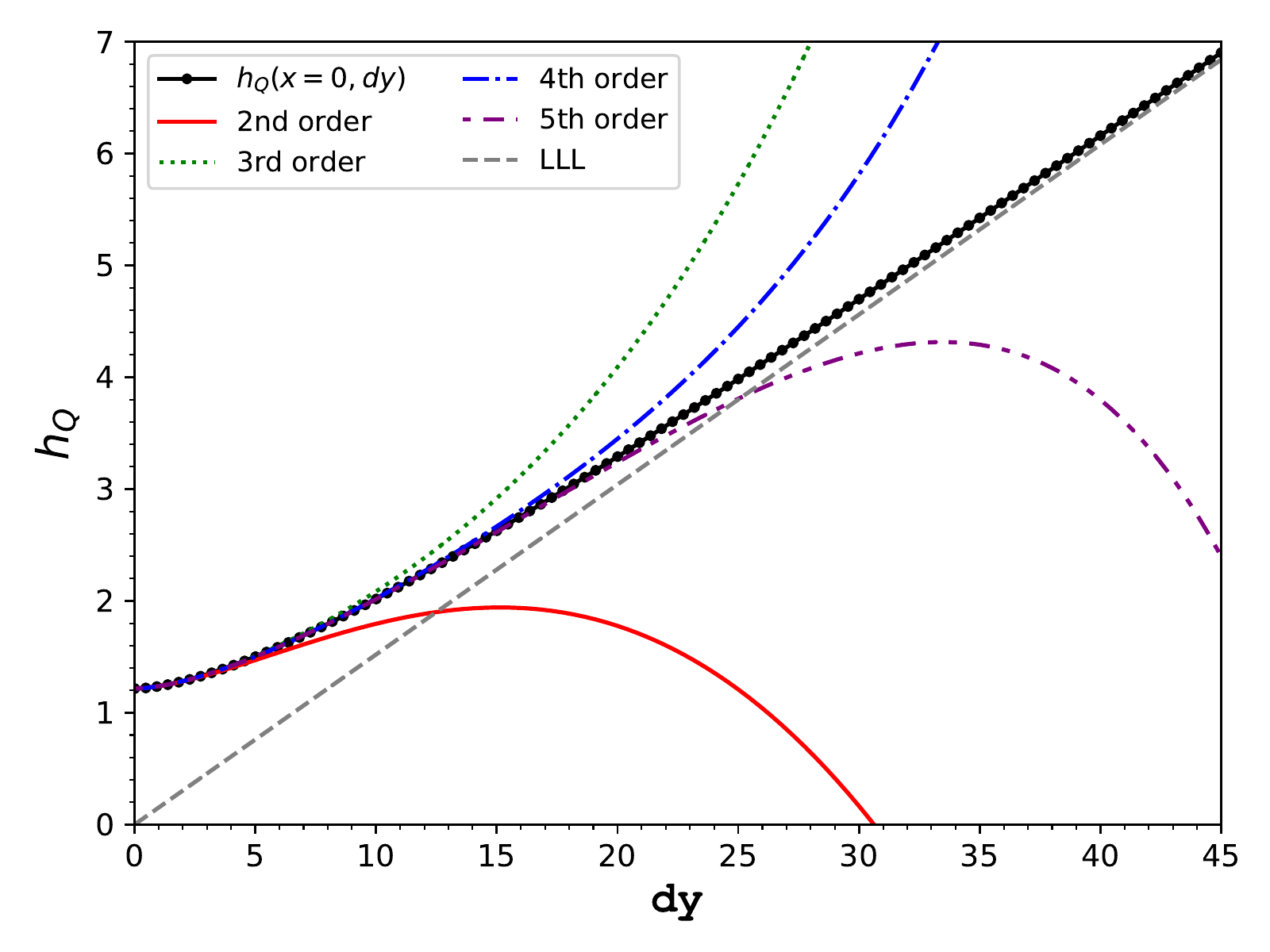}
	\caption{Explicit Z(3) breaking field as a function of $dy = 2 \vert qB \vert /T^2$ at $x = m/T = 2$ (left) and at $x = 0$ (right).
	The robustness of the small $dy$ expansion schemes and the lowest Landau level (LLL) result for $dy \gg 1$ are shown.}
\label{fig11}
\end{figure*}

\subsection{Strong magnetic field}

The strong magnetic field limit ($ dy \gg 1$) is easy to understand: it is dominated by the lowest Landau level (LLL), i.e. the $k=0, \sigma = 0$ part of the sum \eqref{eq:sum}:
\begin{align}
	\begin{split}
		h_Q(x, dy \gg 1) &\rightarrow h_{\rm LLL}(x, dy) \\
		h_{\rm LLL}(x, dy) &= \frac{3}{2 \pi^2} \, x K_1(x) \, dy.
	\end{split}
\end{align}
	Naturally,  $h_{\rm LLL}$ depends linearly on $dy$ (the magnetic field). Contributions from the rest of sum \eqref{eq:sum} are exponentially suppressed. They appear more prominently as $B$ decreases, more so for the case of small $m$ than for large $m$. In fact, the correction terms becomes quadratic in the limit of small $B$, which we turn to next.

\subsection{ weak magnetic field, $x \neq 0$ case }

We shall employ the Riemann Zeta regularization scheme to derive the expansion of $h_Q(x, dy)$ in powers of $dy$.
Noting,  that the sum \eqref{eq:sum} with $\sigma = 0$, compared to $\sigma = 1$, differs only by the LLL term, we write
\begin{align}
	\label{eq:rearrange}
	h_Q^B(x,dy)=\frac{3}{\pi^2} \, S(x, dy) - \frac{3}{2 \pi^2} \, x K_1(x) \, dy
\end{align}
where
\begin{align}
	S(x, dy) = dy \, \sum_{k=0}^\infty \, \sqrt{x^2 + k dy} \, K_1(\sqrt{x^2 + k dy}).
\end{align}

The scheme works by directly handling the $k$-sum in the expansion in powers of $dy$:
\begin{align}
	\begin{split}
		S(x, dy) &= \mathcal{I}(x) + (dy) \times (x K_1(x)) \times [ \sum_{k=0}^{\infty} \, (1) ] \\
		&\quad + (dy)^2 \times (-\frac{1}{2} K_0(x)) \times [ \sum_{k=0}^{\infty} \, k ] \\
		&\quad + (dy)^3 \times (\frac{1}{8 x} K_1(x)) \times [ \sum_{k=0}^{\infty} \, k^2 ] \\
		&\quad + (dy)^4 \times (-\frac{1}{48 x^2} K_2(x)) \times [ \sum_{k=0}^{\infty} \, k^3 ] \\
		&\quad + \cdots.
	\end{split}
\end{align}
The first term is simply given by the Riemann integral:
\begin{align}
	\begin{split}
		\mathcal{I}(x) &= \int_0^\infty  dy \, \sqrt{x^2 + y} \, K_1(\sqrt{x^2 + y}) \\
		&= 2 \, x^2 K_2(x).
	\end{split}
\end{align}
\noindent This recovers the $B \rightarrow 0$ result of $h_Q^B$ in Eq. \eqref{eq:h0}.
A key step of the method is to identify the $\zeta$ function
\begin{align}
	\zeta(s) = \sum_{k=1}^\infty \, \frac{1}{k^s}
\end{align}
\noindent and represent the various divergent sum by the analytic continuation $\zeta(-n)$:
\begin{align}
	\begin{split}
		\sum_{k=0}^{\infty} \, k &\rightarrow \zeta(-1) = -1/12 \\
		\sum_{k=0}^{\infty} \, k^2 &\rightarrow \zeta(-2) = 0 \\
		\sum_{k=0}^{\infty} \, k^3 &\rightarrow \zeta(-3) = 1/120 \\
		\sum_{k=0}^{\infty} \, k^4 &\rightarrow \zeta(-4) = 0 \\
		\sum_{k=0}^{\infty} \, k^5 &\rightarrow \zeta(-5) = -1/252 \\
		\cdots.
	\end{split}
\end{align}
\noindent These results follow from the general formula
\begin{align}
	\zeta(-n) = (-1)^n \, \frac{B_{n+1}}{n+1},
\end{align}
where $B_N$'s are the Bernoulli's numbers. It follows,  that all sums involving positive even powers vanish as the corresponding Bernoulli's numbers are zero.
Note also,  that the $k=0$ term considered so far does not contribute. It does, however, contribute to the sum
\begin{align}
	\sum_{k=0}^{\infty} \, 1 = 1 + \sum_{k=1}^{\infty} \, 1 \rightarrow 1 + \zeta(0) = 1/2.
\end{align}
\noindent Using these results we obtain, up to 6th order:
\begin{align}
	\begin{split}
		S(x, dy) &= \mathcal{I}(x) + \tilde{c}_1 \, (dy) \\
		&\quad + \tilde{c}_2 \, (dy)^2 + \tilde{c}_4 \, (dy)^4 + \tilde{c}_6 \, (dy)^6 + \ldots \\
		\tilde{c}_1 &= \frac{1}{2} \, x K_1(x) \\
		\tilde{c}_2 &= \frac{1}{24} \, K_0(x) \\
		\tilde{c}_4 &= \frac{-1}{5760 x^2} \, K_2(x) \\
		\tilde{c}_6 &= \frac{6\, x K_1(x) + (24+x^2) K_2(x)}{967680 x^6}.
	\end{split}
\end{align}
Note, that the $\tilde{c}_1$ term cancels the LLL term in $h_Q^B$ in Eq. \eqref{eq:rearrange}, giving
\begin{align}
	\label{eq:h1}
	h_Q^B(x,dy) = \frac{3}{\pi^2} \times ( \mathcal{I}(x) + \tilde{c}_2 \, dy^2 + \tilde{c}_4 \, dy^4 + \ldots ).
\end{align}
As discussed in the text, the correction term starts from $dy^2$ order.
\begin{figure*}[t]
\centering
\includegraphics[width=0.49\linewidth]{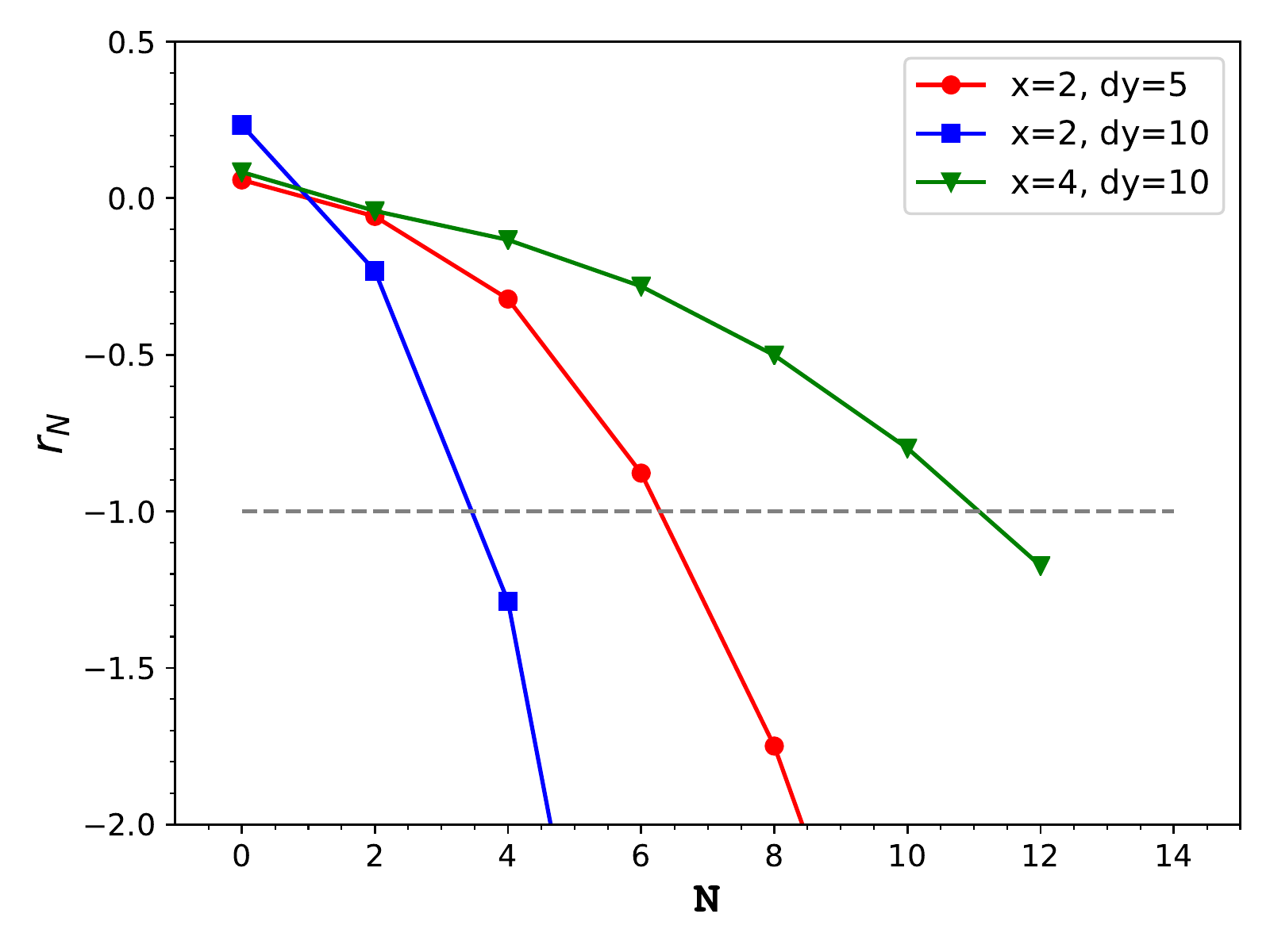}
\includegraphics[width=0.49\linewidth]{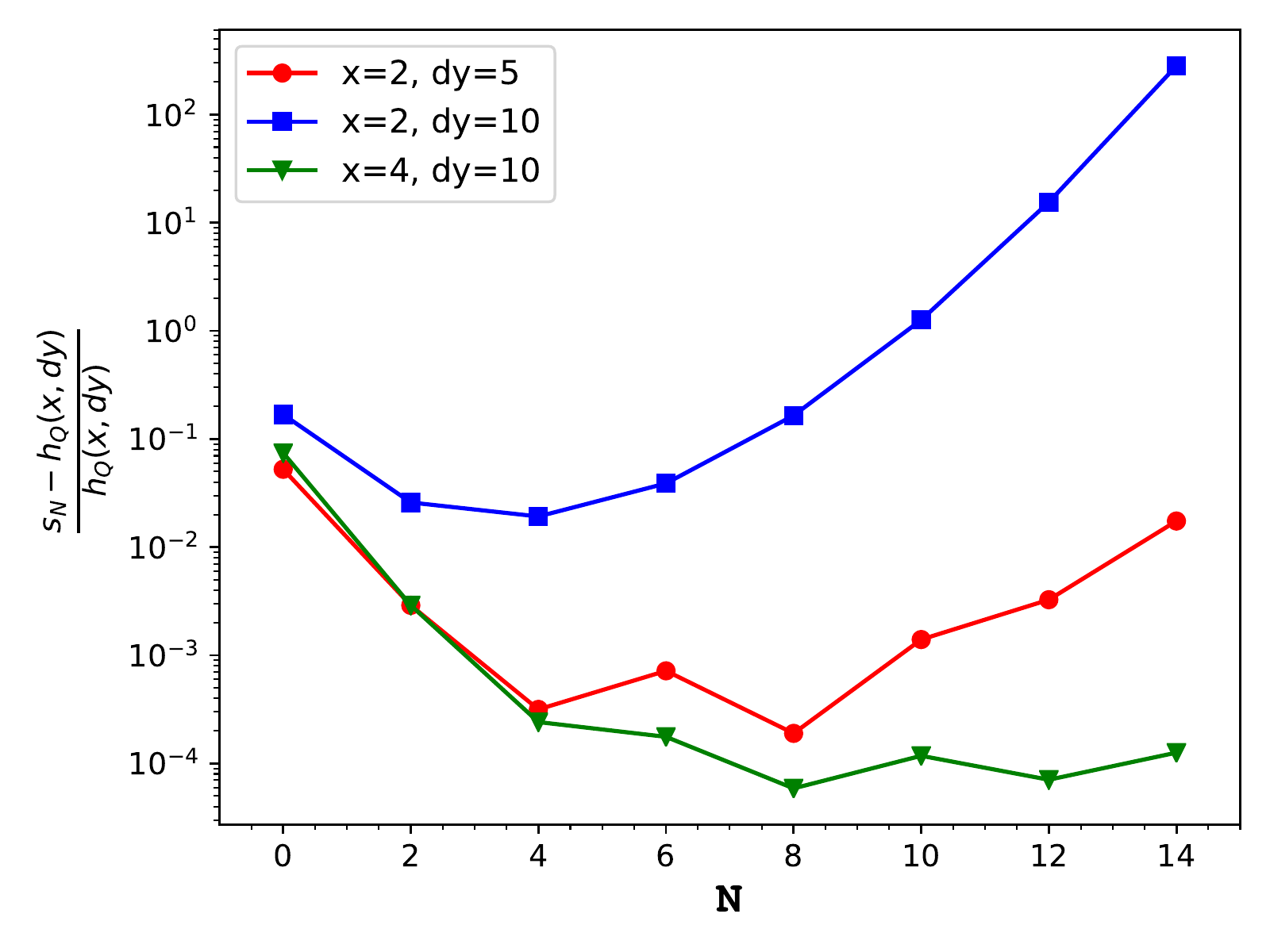}
	\caption{Left: ratio $r_N$ defined in Eq. \eqref{eq:rn} for different values of $(x, dy)$ useful for studying the asymptotic expansion~\eqref{eq:h1}.
	Negative values of $r_N$ suggest that the series is alternating. The optimal number of terms $N^*$ for estimating $h_Q^B(x, dy)$ is given by
	the $N$ at which $\vert r_N \vert $ reaches unity. Right: the relative error of the asymptotic series when summing terms up to the $N$-th order.
	The error first drops and starts to grow again when $N > N^*$.
}
\label{fig12}
\end{figure*}
In Fig.~\ref{fig11} (left) we show the efficacy of the approximation scheme.
Note,  that Eq.~\eqref{eq:h1} is an asymptotic expansion, i.e. the series formally diverges,
and has a zero radius of convergence.
Nevertheless, it can still provide an adequate approximation of the full result using a few terms.
The optimal number of terms to keep ($N^*$) can be inferred by study the ratio $r_N$ as a function of $N$:

\begin{align}
	\label{eq:rn}
	r_N = \frac{\tilde{c}_{N+2}}{\tilde{c}_{N}} \times dy^2.
\end{align}
As the magnitude of the ratio exceeds unity, the accuracy of the sum up to the Nth term, $s_N$, starts to deteriorate.
Figure \ref{fig12} shows the ratios $r_N$ and the corresponding relative errors of $s_N$ for typical values of $(x, dy)$.
The fact that $r_N$'s are negative means that the series is alternating, also evident from Fig. \ref{fig11} (left).
The value of $N^*$ can be extracted where the magnitude of $r_N$ exceeds unity. For example, at $(x = 2, dy = 10)$, the series best approximates the exact result when keeping terms up to $N^*=4$th order.
Including higher order corrections beyond $N^*$ would lead to a worse approximation.
This also explains the failure of the weak field expansion beyond $dy \approx 12$ seen in Fig. \ref{fig11} (left).
Lastly, we note,  that $N^*$ increases with a heavier quark (large $x$) or a weaker magnetic field (small $dy$).
In both cases, the accuracy of the scheme improves significantly,  as shown in Fig. \ref{fig12} (right).
\subsection{ Weak magnetic field, $x=0$ case }

The weak field expansion for the $x=0$ case cannot be simply obtained from the $ x \rightarrow 0 $ limits of the previous results $(\tilde{c}_N(x))$. For example, $\tilde{c}_2 \propto K_0(x) \rightarrow -\infty$ as $x\rightarrow 0$.
Nevertheless, both $h_Q^B(x=0,dy)$ and $S(x=0, dy)$ are finite functions of $dy$.
It turns out,  that there exists a hidden nonanalyticity, hindering the expansion of $S(x=0, dy)$ in powers of $dy$.

To proceed we apply the same Riemann Zeta regularization scheme on $S_0=S(x=0, dy)$:
\begin{align}
	\begin{split}
		S_0 &\simeq dy \, \sum_{k=0}^{\infty} \, \sqrt{k dy} \, K_1(\sqrt{k dy}) \\
		&= \mathcal{I}(x=0) + (dy) \times [ \sum_{k=0}^{\infty} \, (1) ] \\
		&\quad + (dy)^2 \times \left( \frac{-1 + 2 (\gamma_E-\ln 2)}{4} \right) \times [ \sum_{k=0}^{\infty} \, k ] \\
		&\quad + (dy)^2 \times \frac{1}{4} \times [ \sum_{k=0}^{\infty} \, k \ln k ] \\
		&\quad + (dy)^2 \ln (dy) \times \frac{1}{4} \times [ \sum_{k=0}^{\infty} \, k ] + \cdots.
	\end{split}
\end{align}
Here, $\gamma_E \approx 0.577216$ is the Euler's constant.

The first term works out to be $\mathcal{I}(0) = 4$.
For the correction terms we see two new complications:
First, the existence of terms like $(dy)^2 \ln (dy)$, which explains naturally the divergence of $\tilde{c}_2 (x \rightarrow 0)$. Second, the contribution from terms such as
\begin{align}
	\sum_{k=0}^{\infty} \, k \ln k.
\end{align}
\noindent These sums can be related to the derivative of $\zeta(s)$ via
\begin{align}
	\zeta^\prime(s) = (-1) \, \sum_{k=1}^\infty \, \frac{\ln k}{k^s},
\end{align}
\noindent and for our purpose we need the value
\begin{align}
	\sum_{k=0}^{\infty} \, k \ln k &\rightarrow -\zeta^\prime(-1) \approx 0.165421.
\end{align}

Carry out the scheme up to the 5th order, the expression of $S(x=0, dy)$ reads
\begin{align}
	\begin{split}
		S(x=0, dy) &\approx 4 + \frac{1}{2} \, dy \\
		&\quad + ( \tilde{d}_{2a} + \tilde{d}_{2b} \, \ln dy ) \times (dy)^2 \\
		&\quad + d_3 \times (dy)^3 \\
		&\quad + ( \tilde{d}_{4a} + \tilde{d}_{4b} \, \ln dy ) \times (dy)^4 \\
		&\quad + d_5 \times (dy)^5 + \ldots
	\end{split}
\end{align}
where
\begin{align}
	\begin{split}
		\tilde{d}_{2a} &= \frac{1 + 2 (\ln2 - \gamma_E) + 12 (-\zeta^\prime(-1))}{48} \\
		\tilde{d}_{2b} &= -\frac{1}{48} \\
		\tilde{d}_{3} &= \frac{1}{32} \, (-\zeta^\prime(-2)) \\
		\tilde{d}_{4a} &= \frac{(-\zeta(-3)) ( 10 +6 (\ln2 - \gamma_E) ) + 3 (-\zeta^\prime(-3))}{2304} \\
		\tilde{d}_{4b} &= \frac{3}{2304} \, \zeta(-3) \\
		\tilde{d}_{5} &= \frac{1}{36864} \, (-\zeta^\prime(-4)).
	\end{split}
\end{align}

Finally, the Z(3) breaking strength can be obtained via
\begin{align}
	\begin{split}
		h_Q^B(x=0, dy) &=  \frac{3}{\pi^2} \, S(x=0, dy) - \frac{3}{2 \pi^2} \, dy \\
		&\approx \frac{12}{\pi^2} + ( \tilde{d}_{2a} + \tilde{d}_{2b} \, \ln dy ) \times (dy)^2 \\
		&\quad + d_3 \times (dy)^3 \\
		&\quad + ( \tilde{d}_{4a} + \tilde{d}_{4b} \, \ln dy ) \times (dy)^4 \\
		&\quad + d_5 \times (dy)^5 + \ldots
	\end{split}
\end{align}
Again, we observe an explicit cancellation of the linear term, and the correction starts at quadratic order.
The effectiveness of the approximation scheme is shown in Fig. \ref{fig11} (right).
Although the accuracy improves with more and more terms, this is of limited use. For large values of $dy$,
the LLL limit

\begin{align}
	h_{\rm LLL}(x=0, dy) =  \frac{3}{2 \pi^2} dy,
\end{align}
should be used.
\vfill

\section{NJL model at $T=0$, finite $B$}
\label{app2}

In this appendix, we provide further details for the empirical fit function $\mathcal{F}_0(B)$ in Eq.~\eqref{eq:bigparam},
 which describes the increase of the chiral condensate with the magnetic field at $T=0$, i.e. magnetic catalysis. 
A standard NJL model can capture this phenomenon~\cite{Klevansky:1989vi,Andersen:2014xxa,Miransky:2015ava}. 

The NJL potential~\cite{Boomsma:2009yk} at zero temperature, but finite magnetic field, takes the form: (2-flavor $(u,d)$ system, with $N_c=3$ and 3D cutoff scheme)

\begin{align}
	\label{eq:njlpot}
	\begin{split}
		\mathcal{U}_{NJL}(M_u, M_d) &= U_0 + U_1 + U_B \\
		U_0 &= \frac{1}{2} \left( \frac{(M_u-m)^2}{4 G} + \frac{(M_d-m)^2}{4 G} \right) \\
		U_1 &= -2 N_c \sum_{f=u,d} \, \int^\Lambda \frac{d^3 p}{(2 \pi)^3} \, \sqrt{p^2 + M_f^2} \\
		U_B &= -\frac{N_c}{2 \pi^2} \sum_{f=u,d} \, (q_f B)^2 \, \times \\
		& \left( \zeta^\prime(-1, x_f) - \frac{1}{2} \, (x_f^2-x_f) \ln x_f + \frac{x_f^2}{4} \right) \\
		x_f &= \frac{M_f^2}{2 \vert q_f B \vert}. 
	\end{split}
\end{align}
Here we extend the discussion in the main text to include (maximal) splittings between the $(u,d)$-quarks at finite $B$~\cite{Boomsma:2009yk}. The gap equation $ \frac{\partial}{\partial M_f} \mathcal{U}_{NJL} = 0 $ for flavor $f$ can be cast into the following form:

\begin{align}
	\label{eq:gapp}
	\begin{split}
		0 &= R \, (1-\frac{m/\Lambda}{y}) - f(y) - b^2 \, \frac{1}{y} \, \frac{\partial g(x_f(y))}{\partial y} \\
		  f(y) &= \sqrt{1 + y^2} - y^2 \, \sinh^{-1} (1/y) \\
		  g(x_f) &= \zeta^\prime(-1, x_f) - \frac{1}{2} \, (x_f^2-x_f) \ln x_f + \frac{x_f^2}{4} \\
		   x_f &= \frac{y^2}{2 b},
	\end{split}
\end{align}
where $y = \frac{M_f}{\Lambda}$, $G_{crit} = \frac{\pi^2}{6}$, $R = \frac{G_{crit.}}{G \Lambda^2}$, and $b = \frac{\vert q_f B \vert}{\Lambda^2}$. 
 
\begin{figure*}[t]
\centering
\includegraphics[width=0.49\linewidth]{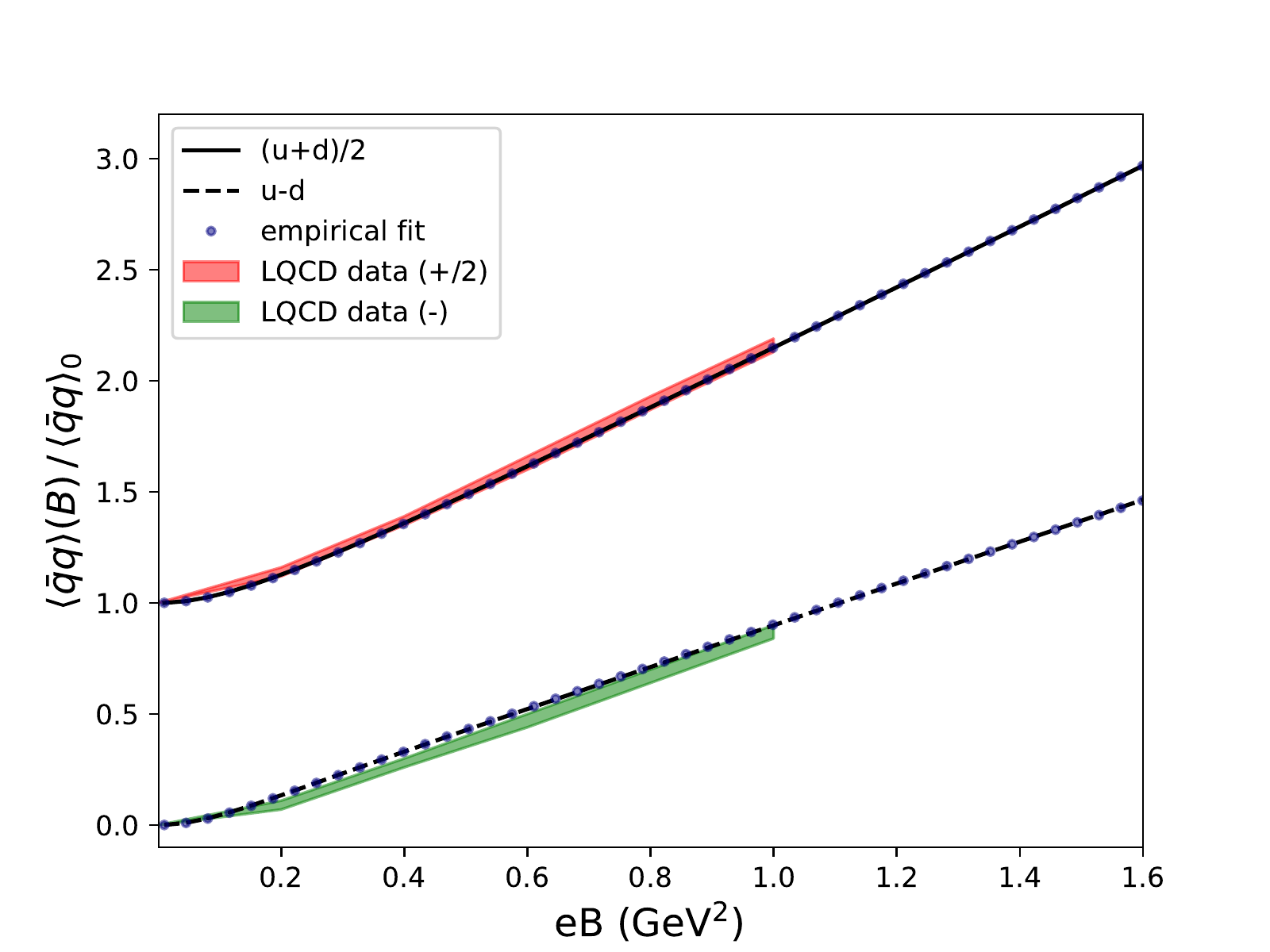}
\includegraphics[width=0.49\linewidth]{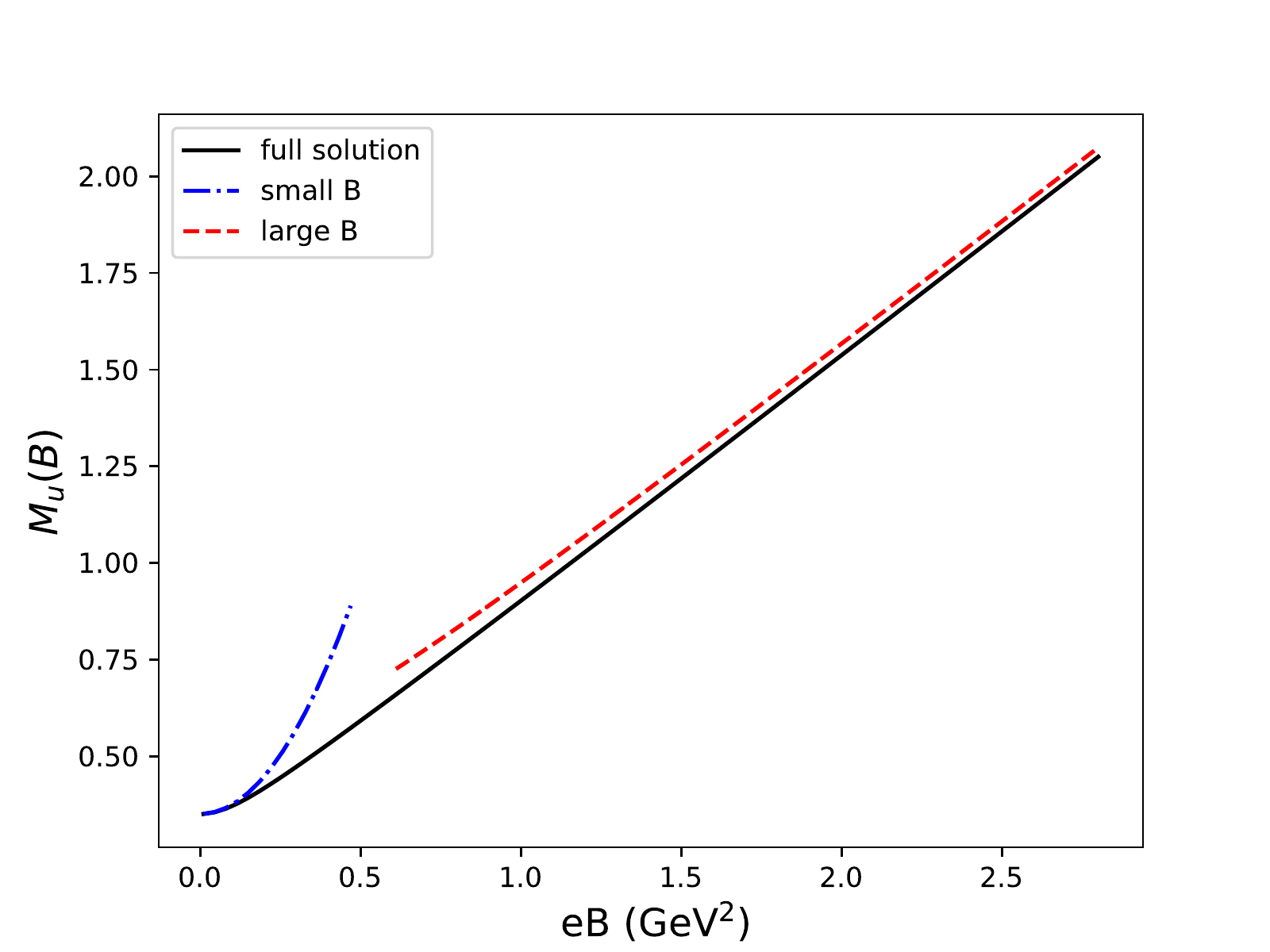}
	\caption{Left: the chiral condensate computed in an NJL model for quarks content: average $(u+d)/2$ and difference $(u-d)$, normalized to the vacuum condensate, versus the strength of the external magnetic field. The LQCD results are extracted from Ref.~\cite{Bali:2012zg}. The empirical fits are based on Eq. \eqref{eq:goodfit}. Right: constituent quark mass function and the asymptotes \eqref{eq:sol1} and \eqref{eq:sol2}.}
\label{fig13}
\end{figure*}

While there is no known exact analytic solution for Eq.~\eqref{eq:gapp}, numerically solving it is straightforward. 
With suitably tuned model parameters: $G \Lambda^2 = 2.435$, $\Lambda = 0.51507$ GeV, 
it can effectively describe the LQCD result~\cite{Bali:2012zg}. 
The $B$-dependent chiral condensates, for the sum and difference of the ($u,d$)-quarks, normalized to the vacuum value, 
is shown in Fig.~\ref{fig13} (left). 

We see that the condensate increases with an external magnetic field: 
rather slowly at first, and turns linear at large fields. 
To understand the latter behavior, we investigate the solution of Eq.~\eqref{eq:gapp} at large $b$. 
We expect $y$ to grow, and it turns out that $x_f = \frac{y^2}{2 b} $ is also not small: $x_f >= 0.4$, justified {\it a posteriori} after the numerical solution. 
Hence, we can take the large argument limit of the functions

\begin{align}
	\label{eq:fglim}
	\begin{split}
		f(y) &\approx \frac{2}{3 y} \\
		g(x_f) &\approx \frac{1}{12} \, \ln x_f,
	\end{split}
\end{align}
and the gap equation becomes a quadratic equation

\begin{align}
		0 &\approx R \, (1-\frac{m/\Lambda}{y}) - \frac{2}{3 \, y} - b^2 \, \frac{1}{6 \, y^2}
\end{align}
with solution
\begin{align}
	\label{eq:sol1}
	\begin{split}
		y &\approx \tilde{y}_0 \times \frac{1 + \sqrt{ 1 + \alpha \, b^2}}{2} \\
		\tilde{y}_0 &= m/\Lambda + \frac{2}{3 \, R} \\
		\alpha &= \frac{\frac{2}{3 \, R}}{(m/\Lambda + \frac{2}{3 \, R})^2}.
	\end{split}
\end{align}
Hence the increase becomes linear at very large $b$'s. 
Note that the $b=0$ limit of Eq.~\eqref{eq:sol1}, i.e. $y_0 \rightarrow \tilde{y}_0 = m/\Lambda + \frac{2}{3 R}$, 
is a solution to the gap equation~\eqref{eq:gapp} at $b=0$ and large coupling $G$, i.e. small $R$.

The small $b$ correction is also easy to understand.
In this case $x_f = \frac{y^2}{2 b} $ is naturally large, 
and the approximate formula~\eqref{eq:fglim} for $g(x_f)$ (but not necessarily for $f(y)$) remains valid. 
By expanding $y$ around $y_0$, the latter solves $R \, (1-\frac{m/\Lambda}{y_0}) = f(y_0)$, we obtain

\begin{align}
	\label{eq:sol2}
	\begin{split}
		0 &\approx R \, (1-\frac{m/\Lambda}{y_0}) - f(y_0) -f^\prime(y_0) \Delta y - b^2 \frac{1}{6 y_0^2} \\
		\implies  y &\approx y_0 - \frac{1}{ 6 \, y_0^2 \, f^\prime(y_0)} \, b^2. 
	\end{split}
\end{align}
Note that $f^\prime(y_0)$ is negative, and indeed we see that the initial correction is positive and quadratic in $B$.
These asymptotes are shown in Fig.~\ref{fig13} (right).

The fit $\mathcal{F}_0(B)$ described in Eq.~\eqref{eq:bigparam} is an empirical fit to this NJL model.
Based on the present analysis, we propose an {\it ansatz} of the following form:

\begin{align}
	\label{eq:goodfit}
	\frac{\langle \bar{q} q \rangle_f(B)-\langle \bar{q} q \rangle_0}{ \langle \bar{q} q \rangle_0} &= a_1 \times (\sqrt{1+a_2 \, \, (q_f B)^2} -1).
\end{align}
The model parameters are given by: $a_1 = 0.257$, $a_2=115.5 \, {\rm GeV}^{-4}$. 
The efficacy of the fit is clearly displayed in Fig.~\ref{fig13} (left).


\begin{thebibliography}{0}%
\makeatletter
\providecommand \@ifxundefined [1]{%
 \@ifx{#1\undefined}
}%
\providecommand \@ifnum [1]{%
 \ifnum #1\expandafter \@firstoftwo
 \else \expandafter \@secondoftwo
 \fi
}%
\providecommand \@ifx [1]{%
 \ifx #1\expandafter \@firstoftwo
 \else \expandafter \@secondoftwo
 \fi
}%
\providecommand \natexlab [1]{#1}%
\providecommand \enquote  [1]{``#1''}%
\providecommand \bibnamefont  [1]{#1}%
\providecommand \bibfnamefont [1]{#1}%
\providecommand \citenamefont [1]{#1}%
\providecommand \href@noop [0]{\@secondoftwo}%
\providecommand \href [0]{\begingroup \@sanitize@url \@href}%
\providecommand \@href[1]{\@@startlink{#1}\@@href}%
\providecommand \@@href[1]{\endgroup#1\@@endlink}%
\providecommand \@sanitize@url [0]{\catcode `\\12\catcode `\$12\catcode
  `\&12\catcode `\#12\catcode `\^12\catcode `\_12\catcode `\%12\relax}%
\providecommand \@@startlink[1]{}%
\providecommand \@@endlink[0]{}%
\providecommand \url  [0]{\begingroup\@sanitize@url \@url }%
\providecommand \@url [1]{\endgroup\@href {#1}{\urlprefix }}%
\providecommand \urlprefix  [0]{URL }%
\providecommand \Eprint [0]{\href }%
\providecommand \doibase [0]{http://dx.doi.org/}%
\providecommand \selectlanguage [0]{\@gobble}%
\providecommand \bibinfo  [0]{\@secondoftwo}%
\providecommand \bibfield  [0]{\@secondoftwo}%
\providecommand \translation [1]{[#1]}%
\providecommand \BibitemOpen [0]{}%
\providecommand \bibitemStop [0]{}%
\providecommand \bibitemNoStop [0]{.\EOS\space}%
\providecommand \EOS [0]{\spacefactor3000\relax}%
\providecommand \BibitemShut  [1]{\csname bibitem#1\endcsname}%
\let\auto@bib@innerbib\@empty
\end{thebibliography}%


\begin{thebibliography}{100}
\bibitem{DElia:2012ems}
  M.~D'Elia,
  Lect.\ Notes Phys.\  {\bf 871}, 181 (2013).

\bibitem{Kharzeev:2012ph}
  D.~E.~Kharzeev, K.~Landsteiner, A.~Schmitt and H.~U.~Yee,
  Lect.\ Notes Phys.\  {\bf 871}, 1 (2013).


\bibitem{Shovkovy:2012zn}
  I.~A.~Shovkovy,
  Lect.\ Notes Phys.\  {\bf 871}, 13 (2013).


\bibitem{Andersen:2014xxa}
  J.~O.~Andersen, W.~R.~Naylor and A.~Tranberg,
  Rev.\ Mod.\ Phys.\  {\bf 88}, 025001 (2016).

\bibitem{Miransky:2015ava}
  V.~A.~Miransky and I.~A.~Shovkovy,
  Phys.\ Rept.\  {\bf 576}, 1 (2015).





\bibitem{Skokov:2009qp}
  V.~Skokov, A.~Y.~Illarionov and V.~Toneev,
  Int.\ J.\ Mod.\ Phys.\ A {\bf 24}, 5925 (2009).

\bibitem{Voronyuk:2011jd}
  V.~Voronyuk, V.~D.~Toneev, W.~Cassing, E.~L.~Bratkovskaya, V.~P.~Konchakovski and S.~A.~Voloshin,
  Phys.\ Rev.\ C {\bf 83}, 054911 (2011).

\bibitem{Bzdak:2011yy}
  A.~Bzdak and V.~Skokov,
  Phys.\ Lett.\ B {\bf 710}, 171 (2012).

\bibitem{Deng:2012pc}
  W.~T.~Deng and X.~G.~Huang,
  Phys.\ Rev.\ C {\bf 85}, 044907 (2012).
		
\bibitem{Tuchin:2013ie}
  K.~Tuchin,
  Adv.\ High Energy Phys.\  {\bf 2013}, 490495 (2013).


\bibitem{Duncan:1992hi}
  R.~C.~Duncan and C.~Thompson,
  Astrophys.\ J.\  {\bf 392}, L9 (1992).

\bibitem{Ferrer:2012wa}
  E.~J.~Ferrer and V.~de la Incera,
  Lect.\ Notes Phys.\  {\bf 871}, 399 (2013).


\bibitem{Grasso:2000wj}
  D.~Grasso and H.~R.~Rubinstein,
  Phys.\ Rept.\  {\bf 348}, 163 (2001).



\bibitem{DElia:2011koc}
  M.~D'Elia and F.~Negro,
  Phys.\ Rev.\ D {\bf 83}, 114028 (2011).

\bibitem{Bali:2011qj}
  G.~S.~Bali, F.~Bruckmann, G.~Endrodi, Z.~Fodor, S.~D.~Katz, S.~Krieg, A.~Schafer and K.~K.~Szabo,
  JHEP {\bf 1202}, 044 (2012).

\bibitem{Bali:2012zg}
  G.~S.~Bali, F.~Bruckmann, G.~Endrodi, Z.~Fodor, S.~D.~Katz and A.~Schafer,
  Phys.\ Rev.\ D {\bf 86}, 071502 (2012).

\bibitem{Bruckmann:2013oba}
  F.~Bruckmann, G.~Endrodi and T.~G.~Kovacs,
  JHEP {\bf 1304}, 112 (2013).

\bibitem{Bornyakov:2013eya}
  V.~G.~Bornyakov, P.~V.~Buividovich, N.~Cundy, O.~A.~Kochetkov and A.~Schäfer,
  Phys.\ Rev.\ D {\bf 90}, 034501 (2014).

\bibitem{Bali:2014kia}
  G.~S.~Bali, F.~Bruckmann, G.~Endrödi, S.~D.~Katz and A.~Schäfer,
  JHEP {\bf 1408}, 177 (2014).


\bibitem{Endrodi:2015oba}
  G.~Endrodi,
  JHEP {\bf 1507}, 173 (2015).


\bibitem{DElia:2018xwo}
  M.~D'Elia, F.~Manigrasso, F.~Negro and F.~Sanfilippo,
  Phys.\ Rev.\ D {\bf 98}, 054509 (2018).

\bibitem{Endrodi:2019zrl}
  G.~Endrodi, M.~Giordano, S.~D.~Katz, T.~G.~Kovács and F.~Pittler,
  JHEP {\bf 1907}, 007 (2019).






\bibitem{Fraga:2012rr}
  E.~S.~Fraga,
  Lect.\ Notes Phys.\  {\bf 871}, 121 (2013).

\bibitem{Fraga:2013ova}
  E.~S.~Fraga, B.~W.~Mintz and J.~Schaffner-Bielich,
  Phys.\ Lett.\ B {\bf 731}, 154 (2014).

\bibitem{Farias:2014eca}
  R.~L.~S.~Farias, K.~P.~Gomes, G.~I.~Krein and M.~B.~Pinto,
  Phys.\ Rev.\ C {\bf 90}, 025203 (2014).

\bibitem{Ferrer:2014qka}
E.~Ferrer, V.~de la Incera and X.~Wen,
Phys. Rev. D \textbf{91},  054006 (2015)



\bibitem{Pagura:2016pwr}
V.~Pagura, D.~Gomez Dumm, S.~Noguera and N.~Scoccola,
Phys. Rev. D \textbf{95}, 034013 (2017). 

\bibitem{Dumm:2018oop}
D.~Gómez Dumm, M.~Izzo Villafañe, S.~Noguera, V.~Pagura and N.~Scoccola,
[arXiv:1805.04597 [hep-ph]].

\bibitem{GomezDumm:2017jij}
D.~Gómez Dumm, M.~Izzo Villafañe and N.~Scoccola,
Phys. Rev. D \textbf{97}, no.3, 034025 (2018).

\bibitem{Braun:2007bx}
  J.~Braun, H.~Gies and J.~M.~Pawlowski,
  Phys.\ Lett.\ B {\bf 684}, 262 (2010).

\bibitem{Schaefer:2007pw}
  B.~J.~Schaefer, J.~M.~Pawlowski and J.~Wambach,
  Phys.\ Rev.\ D {\bf 76}, 074023 (2007).


\bibitem{Reinosa:2014ooa}
  U.~Reinosa, J.~Serreau, M.~Tissier and N.~Wschebor,
  Phys.\ Lett.\ B {\bf 742}, 61 (2015).

\bibitem{Reinosa:2015oua}
  U.~Reinosa, J.~Serreau and M.~Tissier,
  Phys.\ Rev.\ D {\bf 92}, 025021 (2015).

\bibitem{Fukushima:2012qa}
  K.~Fukushima and K.~Kashiwa,
  Phys.\ Lett.\ B {\bf 723}, 360 (2013).

\bibitem{Fukushima:2017csk}
  K.~Fukushima and V.~Skokov,
  Prog.\ Part.\ Nucl.\ Phys.\  {\bf 96}, 154 (2017).



\bibitem{Lo:2013etb}
  P.~M.~Lo, B.~Friman, O.~Kaczmarek, K.~Redlich and C.~Sasaki,
  Phys.\ Rev.\ D {\bf 88}, no. 1, 014506 (2013).

\bibitem{Lo:2013hla}
  P.~M.~Lo, B.~Friman, O.~Kaczmarek, K.~Redlich and C.~Sasaki,
  Phys.\ Rev.\ D {\bf 88}, 074502 (2013).

\bibitem{Bazavov:2016uvm}
  A.~Bazavov, N.~Brambilla, H.-T.~Ding, P.~Petreczky, H.-P.~Schadler, A.~Vairo and J.~H.~Weber,
  Phys.\ Rev.\ D {\bf 93}, no. 11, 114502 (2016).

\bibitem{Lo:2018wdo}
  P.~M.~Lo, M.~Szymański, K.~Redlich and C.~Sasaki,
  Phys.\ Rev.\ D {\bf 97}, no. 11, 114006 (2018).

\bibitem{Meisinger:2001cq}
P.~N.~Meisinger, T.~R.~Miller and M.~C.~Ogilvie,
Phys. Rev. D \textbf{65}, 034009 (2002).

\bibitem{Dumitru:2012fw}
A.~Dumitru, Y.~Guo, Y.~Hidaka, C.~P.~Altes and R.~D.~Pisarski,
Phys. Rev. D \textbf{86}, 105017 (2012).

\bibitem{Dumitru:2013xna}
A.~Dumitru, Y.~Guo and C.~P.~Korthals Altes,
Phys. Rev. D \textbf{89}, 016009 (2014)



\bibitem{Kashiwa:2012wa}
  K.~Kashiwa, R.~D.~Pisarski and V.~V.~Skokov,
  Phys.\ Rev.\ D {\bf 85}, 114029 (2012).

\bibitem{Kashiwa:2013rm}
K.~Kashiwa and R.~D.~Pisarski,
Phys. Rev. D \textbf{87}, 096009 (2013).



\bibitem{Lo:2014vba}
  P.~M.~Lo, B.~Friman and K.~Redlich,
  Phys.\ Rev.\ D {\bf 90}, 074035 (2014).

\bibitem{Bonati:2015dka}
  C.~Bonati, M.~D'Elia and A.~Rucci,
  Phys.\ Rev.\ D {\bf 92}, 054014 (2015).



\bibitem{Bonati:2017uvz}
  C.~Bonati, M.~D'Elia, M.~Mariti, M.~Mesiti, F.~Negro, A.~Rucci and F.~Sanfilippo,
  Phys.\ Rev.\ D {\bf 95}, 074515 (2017).

\bibitem{Rucci:2019hcd}
  A.~Rucci,
  PhD thesis, INFN.
  doi~\url{http://www.infn.it/thesis/thesis_dettaglio.php?tid=13686}




\bibitem{Binder_1987}
K. Binder,
Rep. Prog. Phys. \textbf{50}, 783 (1987).

\bibitem{Kaczmarek:2005gi}
  O.~Kaczmarek and F.~Zantow,
  hep-lat/0506019.



\bibitem{Bilgici:2008qy}
E.~Bilgici, F.~Bruckmann, C.~Gattringer and C.~Hagen,
Phys. Rev. D \textbf{77}, 094007 (2008).

\bibitem{Morita:2011jv}
K.~Morita, V.~Skokov, B.~Friman and K.~Redlich,
Acta Phys. Polon. Supp. \textbf{5}, 803-814 (2012).



\bibitem{whot}
H.~Saito, S.~Ejiri, S.~Aoki, T.~Hatsuda, K.~Kanaya, Y.~Maezawa, H.~Ohno, and
  T.~Umeda,
Phys. \ Rev. \ D {\bf 84}, 054502 (2011).


\bibitem{Fromm:2011qi}
  M.~Fromm, J.~Langelage, S.~Lottini and O.~Philipsen,
  JHEP {\bf 1201}, 042 (2012).

\bibitem{Ejiri:2019csa}
S.~Ejiri \textit{et al.} [WHOT-QCD],
Phys. Rev. D \textbf{101}, 054505 (2020).


\bibitem{Boomsma:2009yk}
  J.~K.~Boomsma and D.~Boer,
  Phys.\ Rev.\ D {\bf 81}, 074005 (2010).

\bibitem{plm1}
  C.~Ratti, M.~A.~Thaler and W.~Weise,
  Phys.\ Rev.\ D {\bf 73}, 014019 (2006).


\bibitem{Klevansky:1989vi}
S.~Klevansky and R.~H.~Lemmer,
Phys. Rev. D \textbf{39}, 3478 (1989).



\end{thebibliography}
\end{document}